\documentclass[12pt,twocolumn]{article}

\usepackage{amsmath,amsfonts,amssymb}
\usepackage{graphicx}
\usepackage{caption}
\graphicspath{{./figures/}}
\usepackage[margin=0.7in]{geometry}
\usepackage[utf8]{inputenc}
\usepackage{float}
\usepackage[
    backend=biber,
    citestyle=science,
    style=chem-rsc,
    maxbibnames=1,
]{biblatex}
\AtBeginBibliography{\footnotesize}
\usepackage{hyperref}
\hypersetup{
    colorlinks = true,
    allcolors=blue,
}
\usepackage{abstract}

\usepackage{titlesec}
\titlespacing\section{0pt}{12pt plus 4pt minus 2pt}{8pt plus 4pt minus 2pt}
\titleformat{\section}[block]{\scshape\filcenter}{}{1em}{}

\renewcommand\thefigure{\textbf{\arabic{figure}}}

\begin{document} 
\twocolumn[{
\begin{center}
\large\textbf{Unusual magnetotransport in twisted bilayer graphene}
\end{center}

\begin{center}
\small{Joe Finney$^{1,2}$, Aaron L. Sharpe$^{2,3}$, Eli J. Fox$^{1,2}$, Connie L. Hsueh$^{2,3}$, Daniel E. Parker$^4$, Matthew Yankowitz$^{5,6}$, Shaowen Chen$^{4,7,8}$, Kenji Watanabe$^{9}$, Takashi Taniguchi$^{10}$, Cory R. Dean$^7$, Ashvin Vishwanath$^4$, Marc Kastner$^{1,2,11}$, David Goldhaber-Gordon$^{1,2,12}$}
\end{center}

\begin{center}
\footnotesize{$^1$\textit{Department of Physics, Stanford University, 382 Via Pueblo Mall, Stanford, CA 94305, USA}}\\
\footnotesize{$^2$\textit{Stanford Institute for Materials and Energy Sciences, SLAC National Accelerator Laboratory, 2575 Sand Hill Road, Menlo Park, California 94025, USA}}\\
\footnotesize{$^3$\textit{Department of Applied Physics, Stanford University, 348 Via Pueblo Mall, Stanford, CA 94305, USA}}\\
\footnotesize{$^4$\textit{Department of Physics, Harvard University, Cambridge, MA 02138, USA}}\\
\footnotesize{$^5$\textit{Department of Physics, University of Washington, Seattle, WA, USA}}\\
\footnotesize{$^6$\textit{Department of Materials Science and Engineering, University of Washington, Seattle, WA, USA}}\\
\footnotesize{$^7$\textit{Department of Physics, Columbia University, New York, NY 10027, USA}}\\
\footnotesize{$^8$\textit{Department of Applied Physics and Applied Mathematics, Columbia University, New York, NY 10027, USA}}\\
\footnotesize{$^9$\textit{Research Center for Functional Materials, National Institute for Materials Science, 1-1 Namiki, Tsukuba 305-0044, Japan}}\\
\footnotesize{$^{10}$\textit{International Center for Materials Nanoarchitectonics, National Institute for Materials Science,  1-1 Namiki, Tsukuba 305-0044, Japan}}\\
\footnotesize{$^{11}$\textit{Department of Physics, Massachusetts Institute of Technology, 77 Massachusetts Avenue, Cambridge, MA 02139, USA}}\\
\footnotesize{$^{12}$To whom correspondence should be addressed; E-mail: \texttt{goldhaber-gordon@stanford.edu}}
\end{center}

\begin{abstract}
We present transport measurements of bilayer graphene with 1.38° interlayer twist and apparent additional alignment to its hexagonal boron nitride cladding. As with other devices with twist angles substantially larger than the magic angle of 1.1°, we do not observe correlated insulating states or band reorganization. However, we do observe several highly unusual behaviors in magnetotransport. For a large range of densities around half filling of the moiré bands, magnetoresistance is large and quadratic. Over these same densities, the magnetoresistance minima corresponding to gaps between Landau levels split and bend as a function of density and field. We reproduce the same splitting and bending behavior in a simple tight-binding model of Hofstadter’s butterfly on a square lattice with anisotropic hopping terms. These features appear to be a generic class of experimental manifestations of Hofstadter’s butterfly and may provide insight into the emergent states of twisted bilayer graphene.
\end{abstract}
}]

\section{Introduction}
\noindent The mesmerizing Hofstadter butterfly spectrum arises when electrons in a two-dimensional periodic potential are immersed in an out-of-plane magnetic field. When the magnetic flux $\Phi$ through a unit cell is a rational multiple $p/q$ of the magnetic flux quantum $\Phi_0=h/e$, each Bloch band splits into $q$ subbands \cite{hofstadterEnergyLevelsWave1976}. The carrier densities corresponding to gaps between these subbands follow straight lines when plotted as a function of normalized density $n/n_s$ and magnetic field \cite{wannierResultNotDependent1978}. Here, $n_s$ is the density of carriers required to fill the (possibly degenerate) Bloch band. These lines can be described by the Diophantine equation $(n/ns)=t(\Phi/\Phi_0)+s$ for integers $s$ and $t$. In experiments, they appear as minima or zeroes in longitudinal resistivity coinciding with Hall conductivity quantized at $\sigma_{xy}=te^2/h$ \cite{stredaQuantisedHallEffect1982,thoulessQuantizedHallConductance1982}. Hofstadter originally studied magnetosubbands emerging from a single Bloch band on a square lattice. In the following decades, other authors considered different lattices \cite{claroMagneticSubbandStructure1979,hasegawaStabilizationFluxStates1990,liTightbindingElectronsTriangular2011}, the effect of anisotropy \cite{barelliMagneticFieldInducedDirectionalLocalization1999,hasegawaStabilizationFluxStates1990,powellDensityWaveStates2019,sunPossibilityQuenchingIntegerquantumHall1991}, next-nearest-neighbor hopping \cite{claroSpectrumTightBinding1981,gvozdikovEnergySpectrumBloch1994,hanCriticalBicriticalProperties1994,hatsugaiEnergySpectrumQuantum1990,ohEnergySpectrumTriangular2000}, interactions \cite{barelliTwoInteractingHofstadter1997,mishraEffectsInteractionHofstadter2016}, density wave states \cite{powellDensityWaveStates2019}, and graphene moirés \cite{bistritzerMoirButterfliesTwisted2011,moonEnergySpectrumQuantum2012}.

It took considerable ingenuity to realize clean systems with unit cells large enough to allow conventional superconducting magnets to reach $\Phi/\Phi_0\sim 1$. The first successful observation of the butterfly in electrical transport measurements was in GaAs/AlGaAs heterostructures with lithographically-defined periodic potentials \cite{albrechtEvidenceHofstadterFractal2001,geislerDetectionLandauBand2004,schlosserLandauSubbandsGenerated1996}. These experiments demonstrated the expected quantized Hall conductance in a few of the largest magnetosubband gaps. In 2013, three groups mapped out the full butterfly spectrum in both density and field in heterostructures based on monolayer \cite{huntMassiveDiracFermions2013,ponomarenkoCloningDiracFermions2013} and bilayer \cite{deanHofstadterButterflyFractal2013} graphene. In all three cases, the authors made use of the 2\% lattice mismatch between their graphene and its encapsulating hexagonal boron nitride (hBN) dielectric. With these layers rotationally aligned, the resulting moiré pattern was large enough in area that gated structures studied in available high-field magnets could simultaneously approach normalized carrier densities and magnetic flux ratios of 1. Later work on hBN-aligned bilayer graphene showed that, likely because of electron-electron interactions, the gaps could also follow lines described by fractional $s$ and $t$ \cite{spantonObservationFractionalChern2018}.

In twisted bilayer graphene (TBG), a slight interlayer rotation creates a similar-scale moiré pattern. Unlike with graphene-hBN moirés, in TBG there is a gap between lowest and next moiré subbands \cite{caoSuperlatticeInducedInsulatingStates2016}. As the twist angle approaches the magic angle of 1.1° the isolated moiré bands become flat \cite{bistritzerMoireBandsTwisted2011,liObservationVanHove2010}, and strong correlations lead to fascinating insulating \cite{caoCorrelatedInsulatorBehaviour2018,caoUnconventionalSuperconductivityMagicangle2018,polshynLargeLinearintemperatureResistivity2019,saitoIndependentSuperconductorsCorrelated2020,sharpeEmergentFerromagnetismThreequarters2019,stepanovCompetingZerofieldChern,yankowitzTuningSuperconductivityTwisted2019,zondinerCascadePhaseTransitions2020}, superconducting \cite{caoUnconventionalSuperconductivityMagicangle2018,polshynLargeLinearintemperatureResistivity2019,saitoIndependentSuperconductorsCorrelated2020,stepanovCompetingZerofieldChern,yankowitzTuningSuperconductivityTwisted2019,zondinerCascadePhaseTransitions2020}, and magnetic \cite{serlinIntrinsicQuantizedAnomalous2019,sharpeEmergentFerromagnetismThreequarters2019,stepanovCompetingZerofieldChern} states. The strong correlations tend to cause moiré subbands within a four-fold degenerate manifold to move relative to each other as one tunes the density, leading to Landau levels that project only toward higher magnitude of density from charge neutrality and integer filling factors \cite{wongCascadeElectronicTransitions2020,zondinerCascadePhaseTransitions2020}. This correlated behavior obscures the single-particle Hofstadter physics that would otherwise be present.

In this work, we present measurements from a TBG device twisted to 1.38° with apparently aligned hBN. When we apply a perpendicular magnetic field, a complicated and beautiful fan diagram emerges. In a broad range of densities on either side of charge neutrality, the device displays large, quadratic magnetoresistance. Within the magnetoresistance regions, each Landau level associated with $\nu=\pm 8, \pm 12, \pm 16, ...$ appears to split into a pair, and these pairs follow complicated paths in field and density, very different from those predicted by the usual Diophantine equation. Phenomenology similar in all qualitative respects appears in measurements on several regions of this same device with similar twist angles, and in a separate device at 1.59° (see Supplementary Materials for details.)

We can reproduce the unusual features of the Landau levels in a simple tight-binding model on a square lattice with anisotropy and a small energetic splitting between two species of fermions. This is at first glance surprising, since that model does not represent the symmetries of the experimental moire structure. We speculate that the unusual LL features we experimentally observe can  generically emerge from spectra of Hofstadter models that include the same ingredients we added to the square lattice model. With further theoretical work it may be possible to use our measurements to gain insight into the underlying Hamiltonian of TBG near the magic angle.

\begin{figure}[t!]
\includegraphics[width=\columnwidth]{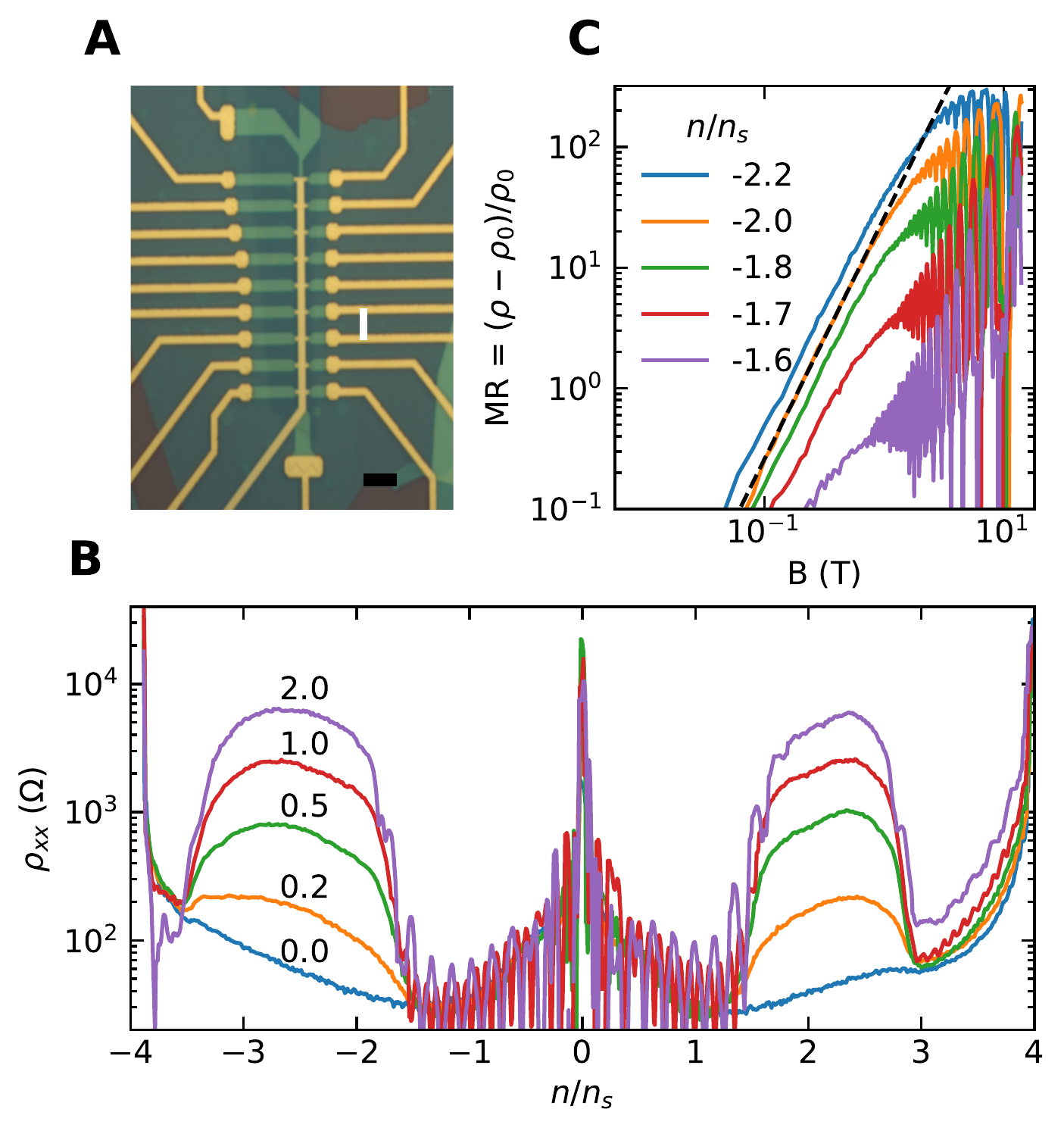}
\captionof{figure}{\textbf{Low-field magnetotransport.} (\textbf{A}) Optical micrograph of the device showing contacts and top gate in gold and hBN in green. We use the large top and bottom contacts to source and drain current. The channel width is 1 \textmu m, and all longitudinal contact pairs are separated by 3 squares. The white line indicates the contact pair that we study throughout this work. Scale bar: 5 \textmu m. (\textbf{B}) Longitudinal resistivity  of the device as density is tuned through empty to full moiré cell at several fixed magnetic fields (in Tesla). The peak at $n=0$ is charge neutrality, and the peaks at the edges of the plot are full filling/emptying of the moiré unit cell. At nonzero fields, there are regions on either side of charge neutrality with large, positive magnetoresistance. (\textbf{C}) Magnetoresistance ratio as a function of field for several fixed densities on a log-log plot. Each trace is offset vertically for clarity. The black dashed line is a quadratic.}
\end{figure}

\begin{figure*}[t!]
\includegraphics[width=\textwidth]{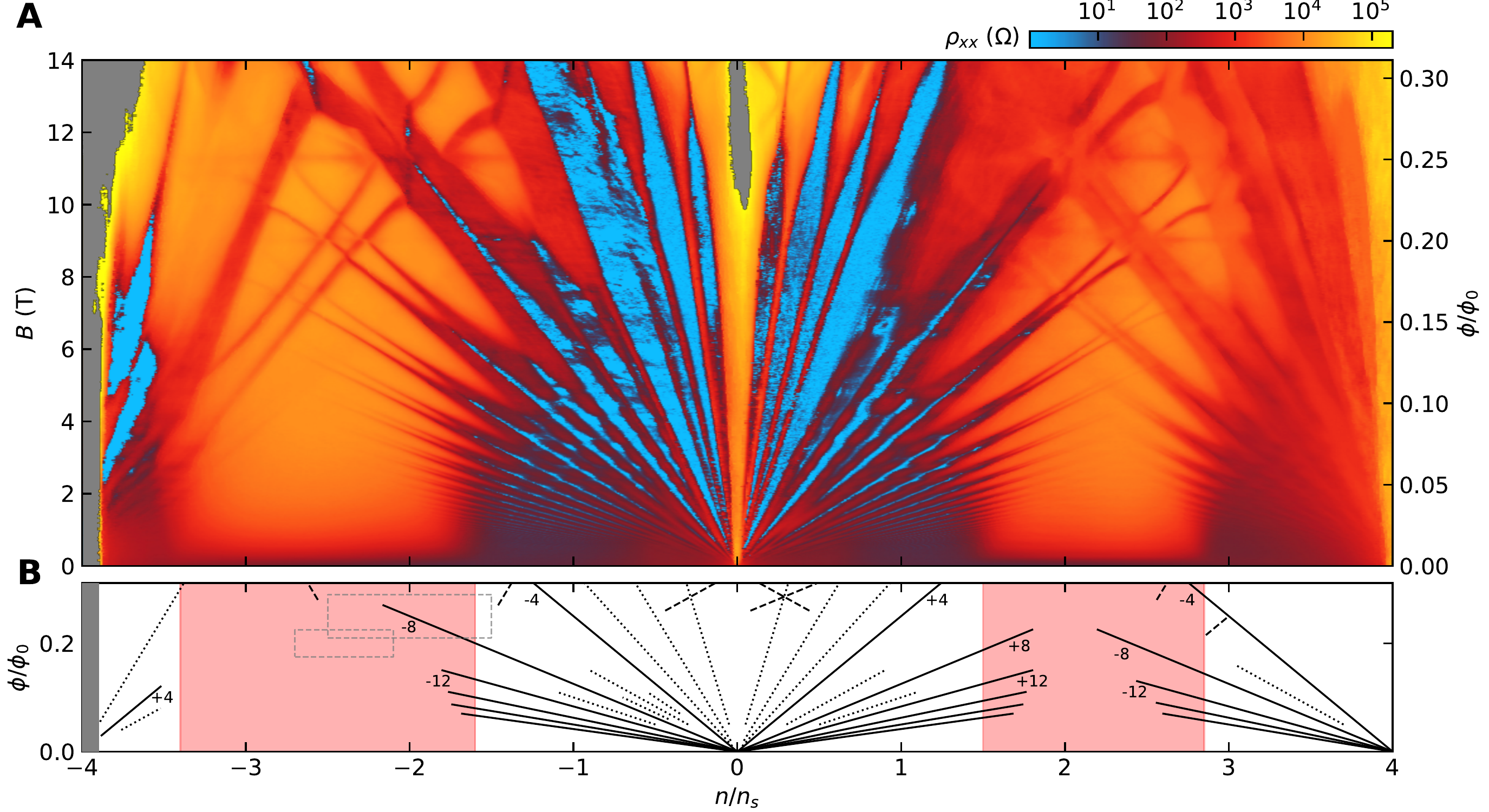}
\captionof{figure}{\textbf{Unusual Landau fan diagram}. (\textbf{A}) Landau fan diagram taken at 26 mK. Landau level gaps are observed as minima in longitudinal resistivity. (\textbf{B}) Schematic fan diagram corresponding to a). Red shaded regions are regions with large magnetoresistance at low field. Solid (dotted) lines are symmetry-preserving (-broken) LLs coming from either charge neutrality or a band edge. Dashed lines are resistance minima corresponding to non-zero $s$ and $t$. The light grey dashed boxes indicate regions reproduced in Fig 3.}
\end{figure*}

\section{Measurements}

\noindent We fabricated this TBG device using the “tear-and-stack” dry transfer method along with standard lithographic techniques \cite{caoSuperlatticeInducedInsulatingStates2016,kimVanWaalsHeterostructures2016}. We encapsulated the device in hBN and included both a graphite back gate and a Ti/Au top gate. When stacking, we attempted to crystallographically align the top layer of graphene to the top layer of hBN. Based on optical micrographs taken during stacking, we appear to have succeeded to within 1°. Using both gates, we could independently tune density $n$ and perpendicular displacement field $D$ \cite{oostingaGateinducedInsulatingState2008}.

Fig. 1a shows an optical micrograph of the completed device: a standard Hall bar with nine voltage probes on each side. The conduction channel is 1 µm wide and all contact pairs are separated by three squares. In this work, we focus on measurements from only one contact pair with twist angle 1.38°±0.01°, however the supplemental material has more information on the other contact pairs. In summary, the twist angle for most contact pairs varies between 1.29° and 1.45°, with the magnetotransport effects that are the focus of this work being peaked around 1.36°. Curiously, two sets of contact pairs near 1.33° display superconductivity (see the supplemental material for details); this is far outside the range of twists around the magic angle where superconductivity has been previously reported for twisted bilayer graphene.

Upon tuning the top gate at fixed magnetic field, we do not observe correlated insulating states at partial fillings of the flat bands (Fig. 1b). This behavior is consistent with reports of samples similarly far above the magic angle \cite{saitoIndependentSuperconductorsCorrelated2020}. Nor do we observe the opening of a gap at charge neutrality or any signatures of ferromagnetism, behaviors which are associated with aligned hBN near the magic angle \cite{serlinIntrinsicQuantizedAnomalous2019,sharpeEmergentFerromagnetismThreequarters2019,stepanovCompetingZerofieldChern}. Instead, in a broad range of densities near half filling, we observe large positive magnetoresistance for both electron and hole doping. The magnetoresistance ratio $[\rho(B)-\rho(0)]/\rho(0)$ is approximately quadratic at low field, reaches over 300, and appears to saturate above 5 T (Fig. 1c).

As we tune both field and density, a complicated series of quantum oscillations originates at the charge neutrality point and $B = 0$ and propagates outwards (Fig. 2a). Near charge neutrality, the Landau levels look similar to those of ordinary magic-angle TBG devices \cite{caoUnconventionalSuperconductivityMagicangle2018,saitoIndependentSuperconductorsCorrelated2020,yankowitzTuningSuperconductivityTwisted2019}, with filling fractions $\nu = \pm 4, \pm 8, \pm 12, ...$ being the most prominent. To within experimental precision, these have zero longitudinal resistance and quantized Hall resistance. As we tune the density into the regions with large magnetoresistance, the Landau levels $\nu = \pm 6, \pm 10, ...$ disappear. Each fourfold degenerate Landau level appears to split into a pair with slopes roughly corresponding to $\nu = \pm 8\pm 0.5, \pm 12\pm 0.5$, and so on (our field range does not allow tracking the $\nu = \pm 4$ levels into the magnetoresistance regions). These split levels do not have zero longitudinal resistance, reaching a minimum of a few hundred ohms. Nor do they follow exactly straight lines. Instead, they bend when approaching other levels. For lack of a better term, we will continue to refer to them as Landau levels.

Landau levels also propagate inward from full filling/emptying of the isolated moiré bands toward lower electron/hole filling, respectively, and these behave similarly to those originating from charge neutrality. We can determine $\Phi/\Phi_0$ by considering the points where these levels cross those originating at charge neutrality. For instance, the level with $\nu = +8$ originating at $n/n_s = -4$ must intersect the level with $\nu = -12$ originating at charge neutrality at $\Phi/\Phi_0=1/5$. In the following discussion, we refer to fields by their values of $\Phi/\Phi_0$.

\begin{figure}[t!]
\includegraphics[width=\columnwidth]{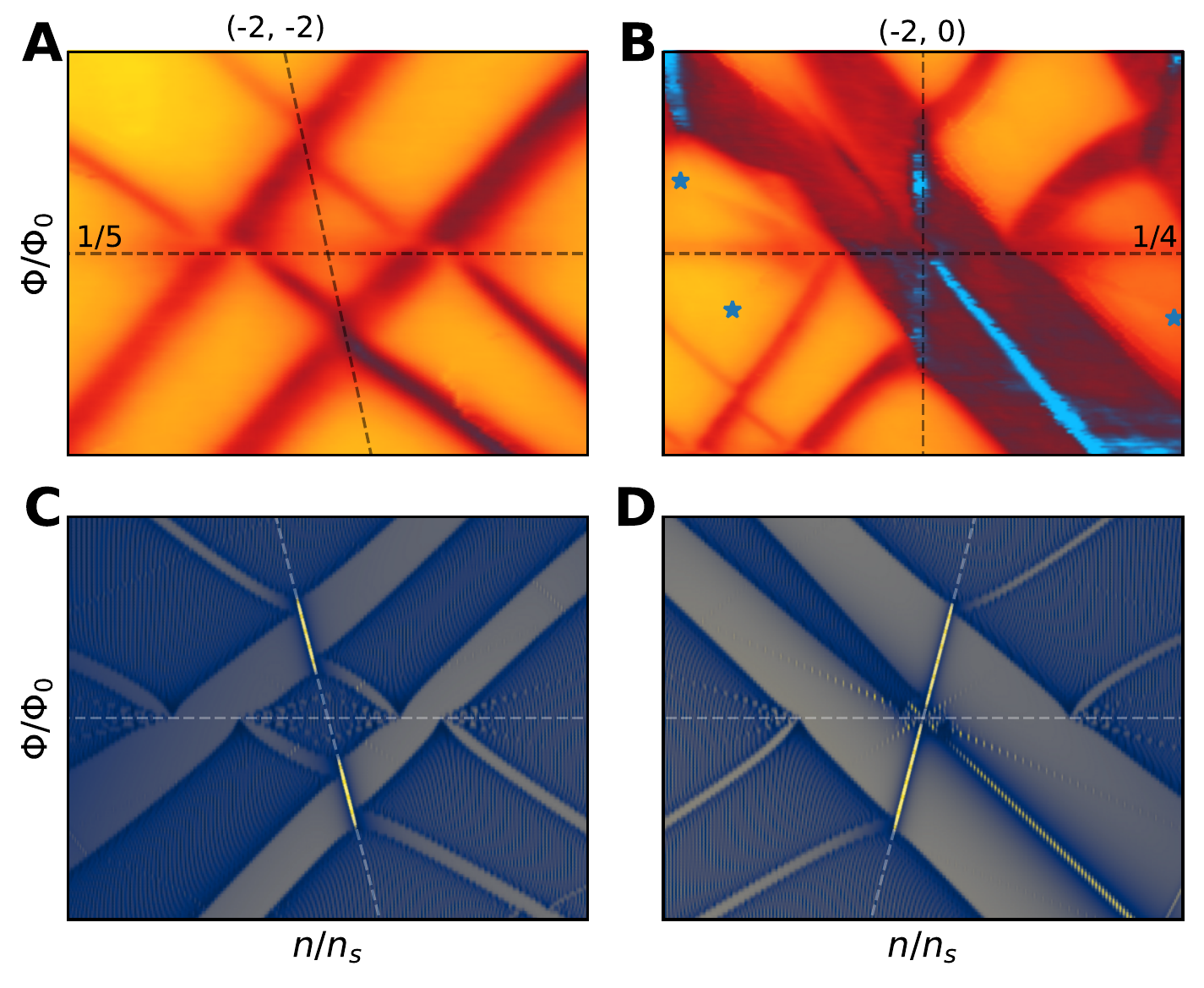}
\captionof{figure}{\textbf{Split Landau level overlap behavior in experiment and computation}. (\textbf{A}) Detail of the crossing of split LLs $-12$ from charge neutrality ($s$, $t$ = $0$, $-12$) and $+8$ from $n/n_s=-4$ ($s$, $t$ = $-4$, $8$), and (\textbf{B}) $-8$ from charge neutrality and $+8$ from $n/n_s=-4$. The horizontal lines are at the indicated $\Phi/\Phi_0$, and the lines with steep slopes are the average ($s$, $t$) of the crossing LLs. For the case of panel A, this is the average of ($0$, $-12$) and ($-4$, $8$) which is ($-2$, $-2$) as indicated. Stars indicate the ends of the faint “extra” LLs originating from the intersections of lower (upper) with upper (lower) split LLs. (\textbf{C}) Computed inverse density of states for $q=1999$ near the crossing of the split levels $s$, $t$ = $-2$, $4$ and $4$, $-6$ for $a_x=1$, $a_y=2$, and $V=0.2$. The nearly vertical dotted line is the average of the two levels, $s$, $t$ = $1$, $-1$, and the horizontal line is $\Phi/\Phi_0= 3/5$. (\textbf{D}) Computed inverse density of states for the crossing of the split levels $s$, $t$ = $0$, $6$ and $2$, $-4$ for $a_x=1$, $a_y=2$, and $V=0.3$. The nearly vertical dotted line is now $s$, $t$ = $1$, $1$, and the horizontal line is $\Phi/\Phi_0= 1/5$. The color scale is as in Fig. 4. The difference in slopes of the crossing diamond shape in model vs. experiment reflects different values of $s$ and $t$ based on a combination of computational convenience and (likely) differences in underlying Hamiltonian.}
\end{figure}

The phenomenology near the intersection of split Landau levels travelling in opposite directions follows a consistent pattern throughout the fan diagram. The example of the $+8$ and $-12$ levels from the previous paragraph is shown in Fig. 3a. As mentioned above, each Landau level splits into a lower and upper level. When a lower (upper) level overlaps with a lower (upper) level moving in the other direction, it changes direction to follow a line originating from half filling ($s = \pm 2$, steeply sloped dashed lines in Fig. 3) with slope equal to the average of the two intersecting levels, which is $-2$ in this case. Within the overlap, the resistivity minima tends to be deeper. Two crossings of a lower with an upper level occur at the same field that the non-split Landau levels would have intersected (horizontal dashed lines in Fig. 3), which is $1/5$ for this example. There is no drop in resistivity where these two intersect. Instead, they appear to displace horizontally by the width of the level that they are crossing before resuming their previous slope. The result of these changes in direction is that in between the overlaps the split LLs are shifted slightly toward each other.

The overlap of the split LLs around $+8$ and $-8$ originating from $n/n_s = -4$ and charge neutrality, respectively, shows the same phenomena (Fig. 3b). In this case, the intersection is at $\Phi/\Phi_0 = 1/4$, and the average slope is $0$, so we see a vertical line of low resistivity. In addition, there are faint additional levels emanating outwards from the two intersections of lower with upper levels.

\begin{figure*}[t!]
\includegraphics[width=\textwidth]{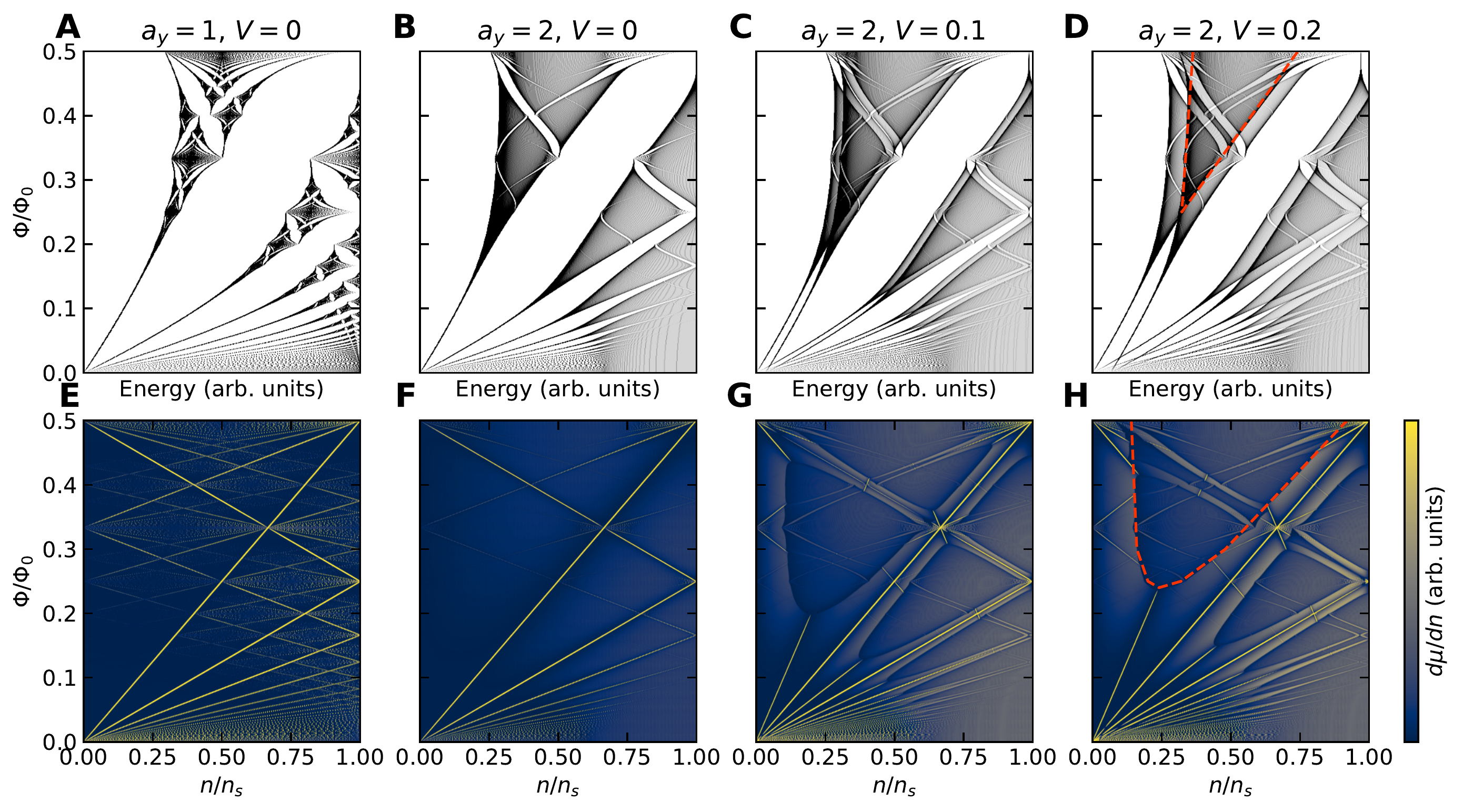}
\captionof{figure}{\textbf{Replication of unusual magnetotransport features in Hofstadter’s butterfly}. (\textbf{A-D}) Energy spectra for the indicated parameters for $q = 1999$, discussed in the main text. Zero energy is the rightmost side of each panel, and they are symmetric about that line. (\textbf{E-H}) Inverse density of states corresponding to spectra in the above panels. The dashed red line bounds one of the high density-of-states region where the two butterflies overlap which correspond to where we see large magnetoresistance in transport.}
\end{figure*}

\section{Discussion}

\noindent Surprisingly, we find that we can reproduce the basic phenomenology observed in the Landau fan diagram with a single-particle calculation that is a simple extension of Hofstadter’s butterfly model. Rather than starting with the standard continuum model of twisted bilayer graphene \cite{bistritzerMoirButterfliesTwisted2011,koshinoEffectiveContinuumModel2020} and attempting to introduce the effects of interactions, we make our calculation in a simple rectangular lattice. We do not expect that the exact details of our calculation match the details in our TBG device, including but not limited to the degeneracies arising from spin and valley. Rather, the replication of several distinctive behaviors in such a different lattice suggests that they are generic features of Hofstadter-like models with anisotropy.

The ordinary Hofstadter butterfly is the result of applying a magnetic field to the tight binding Hamiltonian
\begin{equation}
H = \sum_{\langle i,j \rangle}a_{ij}c_i^\dagger c_j + \mathrm{h.c.}
\end{equation}
where $i,j\in \mathbb{Z}$ index lattice sites, $a_{ij}=a_x=1$ for neighboring sites in the $x$ direction, and $a_{ij}=a_y=1$ for neighboring sites in the $y$ direction. Following many prior works, we numerically solve the associated eigenvalue equation to find the energy spectrum. We then make the simple step of displaying $d\mu /dn$, the inverse density of states \textit{as a function of density}. As we show below, if we slightly modify Hofstadter’s tight-binding Hamiltonian, $d\mu /dn$ emulates the striking phenomenology of our device’s magnetoresistance.

Specifically, we augment the Hamiltonian by allowing the hopping amplitudes to be different in $x$ and $y$ directions ($a_x \neq a_y$) and adding a second fermion species with a tunable energy splitting $V$, yielding
\begin{multline}
H = \sum_{\alpha\in\{A,B\}}\sum_{\langle i,j\rangle}a_{ij}c_{i\alpha}^\dagger c_{j\alpha}\\ + V\sum_i\left(c_{iA}^\dagger c_{iA} - c_{iB}^\dagger c_{iB}\right)+ \mathrm{h.c.}
\end{multline}
In the following text and figures, we set $a_x=1$ and consider only constant $V$. If $V$ is instead set proportional to $B$, to reflect Zeeman splitting of either spins or valleys, the phenomenology is not substantially changed. Fig. 4 shows spectra and the corresponding inverse density of states from the model of Eq. 2 for several values of $a_y$ and $V$.

The spectrum for $a_x=a_y=1$ is identical to the classic isotropic butterfly, and the corresponding inverse density of states demonstrates clear Diophantine behavior (Fig. 4e). However, as there are two fermion species, the Landau levels are doubly degenerate and the gaps follow even-integer slopes only.

Anisotropy has previously been shown to smear out the energy levels and partially close the gaps in the spectrum \cite{barelliMagneticFieldInducedDirectionalLocalization1999,hasegawaStabilizationFluxStates1990,powellDensityWaveStates2019,sunPossibilityQuenchingIntegerquantumHall1991}, which we reproduce by tuning $a_y$ away from $a_x$ (Fig. 4b, f). Upon then introducing a small amount of $V$, a second butterfly pattern appears (Fig. 4c). At low fields and low densities the two butterflies are almost parallel and seldom overlap, and every integer filling of Landau levels gives a ground state with a gap for excitations. However, at higher fields and energies, the anisotropy-broadened butterflies overlap, and odd-integer Landau level fillings have no gap to excitations. The even-integer Landau levels appear to split and bend in the same way as the measured Landau levels in our device, and the behavior at crossings of opposite-polarity Landau levels is also the same as in our device, as shown in the bottom two panels in Fig. 3. The shape of the split LLs is in rough agreement with our experiment for the range of parameters $1.5 < a_y < 3$ and $0.1 < V < 0.3$. The supplemental material shows the behavior of the model outside of this parameter range, and more fully explains the cause of the offsets in split LLs as they intersect.

It is surprising to us that such complex behavior can be reproduced with a simple single-particle model. Nonetheless, several features call for further examination:

First, what causes the striking magnetoresistance? We see very large magnetoresistance in the density-field region of our experimental fan diagram corresponding to where the two broadened butterflies overlap in our model, as indicated in Fig. 4. Though the phenomenological association is clear, it is not obvious to us why overlapping Landau levels should produce such prominent magnetoresistance. One might instead imagine that the magnetoresistance results from coexistence of charge carriers of both signs, since compensated semimetals show some of the strongest known near-quadratic magnetoresistance \cite{aliLargeNonsaturatingMagnetoresistance2014,fatemiMagnetoresistanceQuantumOscillations2017,liangUltrahighMobilityGiant2015}. This phenomenologically-tempting explanation does not simply accord with the persistence of magnetoresistance over a broad gate voltage range (for a fuller discussion, see supplemental material).

Second, what role is played by alignment of the twisted bilayer to hBN? This alignment may modify the single-particle band structure by breaking sublattice symmetry. Nearer the magic angle, this can result in quantum anomalous Hall effect, perhaps particularly when the graphene-graphene and graphene-hBN moiré patterns are commensurate \cite{shiMoirCommensurabilityQuantum2021}. We do not see any features in transport clearly associated with the hBN alignment, so we do not know what role such alignment is playing, if any. In fact, the $\sim 1.5\pm 0.5$° alignment of facets that we observe visually may be between zigzag in one material and armchair in the other, in which case the effect of the hBN on the graphene electronic structure may be much weaker.

Third, does the apparent anisotropic effective Hamiltonian emerge from electron interactions, from uniaxial strain alone, or from some combination such as electronic ordering with order parameter set by strain? Some previous theoretical \cite{chichinadzeNematicSuperconductivityTwisted2020,fernandesNematicityTwistRotational2020,kangNonAbelianDiracNode2020,liuNematicTopologicalSemimetal2021,samajdarElectricfieldtunableElectronicNematic2021,parkerStraininducedQuantumPhase2020} and experimental \cite{choiElectronicCorrelationsTwisted2019,jiangChargeOrderBroken2019,kerelskyMaximizedElectronInteractions2019,rubio-verduUniversalMoirNematic2020} results suggest nematic order at a variety of filling factors within the lowest energy moire miniband manifold, both in TBG relatively close to the magic angle of 1.1°, and in twisted double bilayer graphene. Emergence of nematic order likely heralds an anisotropic effective Hamiltonian, plausibly explaining the correspondence between our device’s behavior and that of our simple model in which anisotropy was built in. We hope our results prompt examination of whether and how such ordering can emerge even far from the magic angle, as in our sample. 

Landau fan diagrams have been a staple of electrical transport measurements for decades, because they give clear insight into the spectrum of electronic states and their filling. In this work, we have identified an entirely new confluence of phenomena in the fan diagram of a TBG device and have found, to our surprise, that this same combination emerges naturally from a single-particle Hamiltonian with anisotropic tunneling.

\setlength\bibitemsep{0pt}
\printbibliography

\noindent\textbf{Acknowledgments}: This work greatly benefited from the advice of M. Zaletel and A. MacDonald, along with ideas from the many scientists with whom we shared these measurements over the last year, including P. Jarillo-Herrero, S. Kivelson, A. Young, B. Feldman, A. Pasupathy, T. Senthil, and A. Mackenzie. \textbf{Funding}: Device fabrication, measurements, and analysis were supported by the U.S. Department of Energy, Office of Science, Basic Energy Sciences, Materials Sciences and Engineering Division, under contract DE-AC02-76SF00515. Measurement infrastructure was funded in part by the Gordon and Betty Moore Foundation’s EPiQS Initiative through grant GBMF3429 and grant GBMF9460, and D.E.P. was supported in part by grant GBMF8683. D.G.-G. gratefully acknowledges support from the Ross M. Brown Family Foundation. Part of this work was performed at the Stanford Nano Shared Facilities (SNSF), supported by the National Science Foundation under award ECCS-1542152. S.C., M.Y., and C.R.D. were supported as part of Programmable Quantum Materials, an Energy Frontier Research Center funded by the U.S. Department of Energy (DOE), Office of Science, Basic Energy Sciences (BES), under award DE-SC0019443. K.W. and T.T. acknowledge support from the Elemental Strategy Initiative conducted by the MEXT, Japan, Grant Number JPMXP0112101001 and JSPS KAKENHI Grant Number JP20H00354. \textbf{Author contributions}: J.F. and C.H. fabricated the device. J.F., A.L.S., and E.J.F. performed transport measurements and analysis. D.E.P. and A.V. contributed the two-species toy model. M.Y., S.C., and C.R.D. fabricated and measured samples D34 and D25. K.W. and T.T. generously supplied the hBN crystals. M.A.K. and D.G.-G. supervised the experiments and analysis. The manuscript was prepared by J.F. with input from all authors. \textbf{Competing interests}: M.A.K. is chair of the Department of Energy Basic Energy Sciences Advisory Committee. Basic Energy Sciences provided funding for this work. \textbf{Data and materials availability}: The data from this study along with all code used to perform analysis and simulation are available at \texttt{https://github.com/spxtr/noblehierarch}.

\onecolumn
\begin{center}
\large\textbf{Supplementary Material}
\end{center}

\section{Fabrication}
We assembled our device using a “tear-and-stack” method. We first prepared a Poly(Bisphenol A carbonate) film stretched over a gel (Gel-Pak DGL-17-X8) and affixed it to a glass slide with double-sided tape. To start stacking, we picked up the top layer of hBN at 80℃. We then used the edge of the hBN flake to pick up and tear the graphene at room temperature. The lower temperature compared to the other steps helps to prevent a common cause of stacking failure for us: graphene outside the region directly contacted by the hBN being picked up or dragged. In this step, we attempted to optically align a long, straight edge of the hBN to a similar edge of the graphene. We then rotated the remaining portion of the graphene flake by 1.2°, picked it up at 80℃, picked up the bottom hBN at 80℃, and then finally picked up a flake of few-layer graphite at 80℃  to form the back gate. We transferred the final stack at 150℃ onto 300-nm-thick SiO$_2$ on degenerately doped Si with pre-patterned alignment marks. The resulting heterostructure is shown in Fig. S1a.

We then used several iterations of standard e-beam lithography to define the Hall bar. We deposited a Ti/Au top gate, etched the Hall bar region using CHF$_3$/O$_2$ (50/5 sccm), and then deposited Cr/Au edge contacts. The device geometry is labelled in Fig. S1b.

\section{Measurements}
All measurements in the main text were taken in a dilution refrigerator with a base temperature of 26 mK at the mixing chamber. The measurement lines include low-pass RF and discrete RC filters at the mixing chamber stage. We used a Stanford Research SR830 lock-in amplifier with a 1 GΩ bias resistor to source an alternating current of 1 nA at roughly 1 Hz. We measured differential voltage pairs with NF Corporation LI-75A voltage preamplifiers and SR830 lock-in amplifiers. We applied gate voltages using Yokogawa 7651 DC voltage sources. We held the Si back gate at a constant 30 V for all measurements to promote transparent contacts.

\section{Twist angle determination}
We initially tested the device in a variable temperature insert (VTI) system at ~1.7 K using a homemade set of lock-in amplifiers that allowed us to measure every longitudinal and Hall voltage pair simultaneously. By sweeping density and field, we can see the overlap of the Landau levels originating at charge neutrality and full-filling, respectively. The spacing of these overlap points is constant in $1/B$ and is directly related to the area of the moiré unit cell, which allows for accurate calculation of the twist angle \cite{caoCorrelatedInsulatorBehaviour2018}. In Fig S1c we show the twist angle variation across the device. The topmost contact pairs have a dramatically different twist angle than the rest. The rest of the contacts vary between 1.28° and 1.45°, with the majority near 1.36°. The contacts near 1.36° display the strongest magnetoresistance. All contact pairs with $\mathrm{MR} > 10$ show the unusual LL behavior discussed in the main text to some degree.

\section{Effect of displacement field}
The Si back gate is screened by the graphite back gate which extends only over the main conduction channel, but is not screened near the contacts. This allows us to set the carrier density near contacts to be high to encourage contact transparency. Meanwhile, the top and back gates allow us to individually tune density and displacement field. Unfortunately, when the back gate is near 0 V we find that our contact resistances dramatically increase in a magnetic field, leading to a loss of signal at lower temperatures. This happens regardless of the Si back gate voltage. Thus, for the measurements in the main text, we fix the back gate at 1.5 V and then sweep the top gate. This means that the measurements are not performed at a constant displacement field as we would prefer. Instead, they follow the black dashed line in Fig. S2c. As we will next discuss, varying the displacement field does not substantially change the phenomenology presented in the main text.

We calculate the displacement field as in reference \cite{sharpeEmergentFerromagnetismThreequarters2019}. Fig. S2 shows the displacement field dependence of the contact pair from the main text at a few fixed magnetic field values. There is no apparent effect at 0 T. At 3 T, negative displacement field enhances the peak value of the magnetoresistance and widens the density extent of the magnetoresistance region. The resistivity at the charge neutrality point increases with increasing magnitude of displacement field, regardless of the polarity of this field. In Fig. S2c, we can clearly observe the split LLs within the magnetoresistance regions at 8 T as pairs of vertical lines, corresponding to constant density, independent of displacement field.

\section{Behavior of other contact pairs}
All measurements in the main text are of contact pair 16 - 17. We present Landau fan diagrams and gate maps of two other contact pairs in Fig. S3. In both cases, the basic phenomenology presented in the main text is reproduced. Contact pair 7 - 8 is very similar to 16 - 17. Contact pair 4 - 5 displays split LLs, though less clearly defined.

In both of these cases, there are oscillations as we tune the displacement field at certain densities. We do not have a satisfactory explanation for this observation, however we note that they appear to be more closely related to the value of the back gate voltage rather than the displacement field (this is reflected in their downward slope in Fig. S3d; lines of constant gate voltage are sloped since a transformation has been applied to make the axes of that figure n and D). Just as the back gate gates the moire channel, the moire channel gates the back gate, so band filling in the 4-5 layer thick back gate is determined primarily by applied back gate voltage. Empirically, at zero back gate voltage, contacts to the moire channel become high resistance, perhaps reflecting low density of carriers in the back gate. At high magnetic field as in Fig. S3c, carrier density in the back gate might be substantially and nonmonotonically modulated with back gate voltage. One way this could affect electronic properties of the moire channel is through changes in screening, given that the lower hBN layer is only about 13 nm thick, comparable with the moiré wavelength. Regardless of the reason for the back-gate-specific effect, the choice of fixing back gate voltage and sweeping top gate for the figure in the main text may actually be superior to holding the displacement field fixed, since to do that would require us to vary the back gate at the same time.

\section{Additional phenomenology near charge neutrality}
Fig. S4 shows the longitudinal magnetotransport at low field and low density for contact pair 7 - 8. The other two contact pairs behave in a similar manner. Some of the Landau level gaps disappear and then reappear as the field is increased. This same behavior can be seen in the model, in Fig. 4g and h from the main text, which is reproduced in S4b: there we can see the Landau levels from the two butterflies intersect at low field, leading to a gradual disappearance and then reappearance of the gap.

Unusual horizontal lines (constant field) appear in between many of the Landau levels. These lines appear to take steps upon crossing Landau levels. These are seen to some extent in all three contact pairs, and are also seen near full-filling and emptying (n/ns = ±4). Within a simple Hofstadter model with only one fermion species but additionally with a next-neighbor hopping term, these horizontal lines can be qualitatively reproduced (Fig S4c and d):  B-field-periodic modulation in the width of the Landau levels causes horizontal lines of reduced resistivity corresponding to fields where the Landau level is narrower and thus has increased DOS. The parameter tuning that gives such modulation is independent of that needed for the phenomenology highlighted in the main text (one can get either, both, or neither). In a more physically realistic model of the moire this phenomenon may be more firmly linked to the rest.

\section{Superconductivity at 1.33°}
Contact pairs 13 - 14 ($\theta$ = 1.33°) and to a lesser extent 3 - 4 ($\theta$ = 1.35°) display evidence of superconductivity near half filling of holes. We show measurements from 13 - 14 in Fig. S5. To our knowledge, superconductivity has not been previously reported for a sample so far above the magic angle, and in our measurements the superconductivity is significantly weaker than that in samples near the magic angle, as demonstrated by its low critical temperature of ~150 mK, low critical field of ~3 mT, and low critical current of ~12 nA.

\section{Low-field Hall effect}
Fig. S6 shows the Hall density for contact pair 13 - 3 ($\theta = 1.33°$). We do not observe a resetting of the Hall density at integer filling factors, as is seen in samples near the magic angle. Instead, roughly in the center of each magnetoresistance region, the Hall density diverges and then changes sign. All contact pairs in the device display similar behavior, as does D34 in the supplement. In these magnetoresistance regions, the Hall slope becomes nonlinear at higher fields. The Hall density shown here is only based on the low-field slope.

\section{Temperature dependence of the magnetoresistance}
Fig. S7 shows the behavior of the magnetoresistance as a function of temperature, qualitatively similar to that seen in WTe$_2$ \cite{aliLargeNonsaturatingMagnetoresistance2014,fatemiMagnetoresistanceQuantumOscillations2017}, Cd$_3$As$_2$ \cite{liangUltrahighMobilityGiant2015}, and other compensated semimetals. In these materials, it is understood that the Hall voltage from one carrier type cancels that from the other type, leading to circular charge carrier trajectories and thus reduced carrier diffusion and positive magnetoresistance.

If the magnetoresistance in our device were the result of a compensated semimetal, we would not expect the magnetoresistance to be so consistent over such a large range of gate-tuned total density. Also, we would expect to see sets of Landau levels originating at the edges of each isolated pocket in the Fermi surface, whereas we only see them originating at charge neutrality and moiré band edges. The split Landau levels have the same periodicity in $1/B$ as each other (Fig. S8), so they correspond to the same density offset from their respective band edge. This has motivated us to seek an alternative explanation.

\section{Hofstadter calculation details}
We start with the Hamiltonian from Eq. 2 of the main text:
\begin{equation}
H = \sum_{\alpha\in\{A,B\}}\sum_{\langle i,j \rangle}a_{ij}c_{i\alpha}^\dagger c_{j\alpha} + V\sum_i\left(c_{iA}^\dagger c_{iA} - c_{iB}^\dagger c_{iB}\right)+ \mathrm{h.c.}
\end{equation}
where $i$ and $j$ are lattice sites of a square lattice with unit lattice constant, and $A$ and $B$ are the two fermion species. In the Landau gauge, the vector potential $\mathbf{A}=(0,Bx,0)$, and the hopping terms in $y$ pick up a Peierl’s phase of $eBx/\hbar=2\pi \phi x$ for $\phi=eB/2\pi\hbar$. Considering only rational values $\phi=p/q$ for coprime $p$ and $q$, the Peierl’s phase repeats after $q$ hops in the $x$ direction, and so we define the magnetic Brillouin zone (MBZ, not to be confused with mini or moiré BZ) as $0<k_x<2\pi/q$ and $0<k_y<2\pi$.

The Peierl’s substitution leads to $c_n^\dagger c_m \rightarrow e^{2\pi i\phi m}c_m^\dagger c_m$, where site $n$ is directly above site $m$ in the $y$ direction, and removes all hopping terms in the $y$ direction from the Hamiltonian, so eigenstates should be plane waves in $y$, which leads to the Hamiltonian
\begin{multline*}
H = \sum_{\alpha\in\{A,B\}}\sum_{m=1}^q\left[ a_y\cos (2\pi \phi m-k_y) c_{m\alpha}^\dagger c_{m\alpha} + (a_xe^{-ik_x}c^\dagger_{m+1,\alpha}c_{m,\alpha} + \mathrm{h.c.})\right]\\ + V\sum_{m=1}^q\left(c_{mA}^\dagger c_{mA} - c_{mB}^\dagger c_{mB}\right)+ \mathrm{h.c.}
\end{multline*}
As this is a finite 1D Hamiltonian, it is now straightforward to numerically compute its eigenvalues. The most convenient way to compute the spectrum over a large range of magnetic field values is to set $q$ to some large prime number (typically thousands, depending on the available computation power) and then solve the above equation for $p$ from 1 to $q - 1$. In principle, we should vary $k_x$ and $k_y$ throughout the MBZ to extract the full energy spectrum. For such large values of $q$, the width of each subband becomes extremely small such that the spectra are effectively constant throughout the MBZ, so we only solve for $k_x=k_y=0$.

As the gaps in the spectrum appear as extremely fine, bright lines in inverse density of states, they are susceptible to aliasing artifacts when the images are downsampled. To suppress these artifacts, we apply a fine gaussian filter ($\sigma \leq 1/q$) to these plots. This does not affect the phenomenology.

We show the behavior of the split LLs for a range of hopping parameters in Fig. S9. We have also made calculations on a hexagonal lattice with anisotropy and two fermion species, and have verified that the behavior is qualitatively similar to that on the square lattice with anisotropy and a second fermion species. Without fine tuning we can replicate the disappearance of odd LLs, the split even LLs, and the split intersection behavior. We also checked that the same behaviors replicate for $V \propto B$ (Fig. S10). Though these are not microscopically faithful models for TBG, many features of the Hofstadter problem are set by the topology of the bands alone and thus should be only weakly model-dependent.

\section{Split Landau level intersection behavior in the model}
The complicated split LL intersection behavior from our model is not difficult to understand, however it is somewhat intricate. We begin by noting that many of the gaps in Hofstadter’s butterfly take horizontal steps\textemdash a discontinuity in energy at a given magnetic field\textemdash when they cross gaps moving the other direction (Fig. S11 a and b). When the second butterfly appears out of the first (Fig. S11c and d), it is shifted horizontally in energy. We then effectively have two Xs next to each other. When two upper or two lower split LLs overlap, the system is simultaneously in the gap of both butterflies, and so we see the average $(s, t)$ of the two intersections. When an upper meets a lower split LL, we are seeing the intersection of LLs within a single butterfly. Therefore, it happens at the appropriate field for that intersection. As there is a background DOS from the other butterfly, we see the horizontal step in energy that is present even in undoubled lattices.

\section{Comparison to additional devices}
We present measurements from two additional devices with twist angles 1.59° (D34, Fig. S12) and 1.52° (D25, Fig. S13). Unlike the device from the main text, both devices have dual graphite gates, and neither device has purposefully-aligned hBN. They were measured in a four-probe configuration at 300 mK with a 10 nA AC current bias at 17.7 Hz. For more details, see ref. 31 and its supplementary text.

Both devices show the same generic Landau level progression as the device from the main text, with dominant Landau levels $\nu = \pm 4, \pm 8, \pm 12, …$. Both devices have regions of  positive magnetoresistance, reaching ratios of $\sim 30$ in D34 and $\sim 10$ in D25 at 3 T. The magnetoresistance is significantly weaker than that of the device in the main text, and it is also not a clean quadratic at low field. While D25 does have an intricate and beautiful fan diagram, it does not display split Landau levels. D34 shows Landau level splitting comparable to that in the device from the main text for filling fractions $\nu = \pm 12, \pm 16, \pm 20, …$. Although much of the fine detail of the split Landau levels appears washed out, the intersecting behavior from the text is also visible for some pairs of overlapping levels.

We summarize the magnetotransport phenomenology for the three devices and their separate contact pairs in Table S1.

\renewcommand{\tablename}{\textbf{Table}}
\renewcommand\thetable{\textbf{S\arabic{table}}}

\begin{table*}[t]
\centering
\begin{tabular}{c c r r r r c}
Device & Contact pair & T (mK) & $\theta$ (°) & $B_{sat}$ (T) & MR$_{max}$ & Split LLs? \\
\hline
Main text & 16 - 17 & 26 & 1.38 & 8 & 340 & Very clear\\
Main text & 7 - 8 & 26 & 1.37 & 8 & 280 & Very clear \\
Main text & 4 - 5 & 26 & 1.35 & 5 & 210 & Clear \\
D34 & & 300 & 1.59 & 3 & 30 & Clear \\
D25 & Holes & 300 & 1.52 & 3 & 10 & Absent \\
D25 & Electrons & 300 & 1.52 & 3 & 10 & Absent \\
\end{tabular}
\caption{\textbf{Summary of the separate contact pairs and devices presented in the text.} The columns are device and contact pair, temperature at which the measurements were taken, twist angle, saturation field, largest magnetoresistance ratio, and whether or not we see clear split Landau levels. Though measurements were made at different temperatures for different devices, all those temperatures were well below 1K, whereas MR is not strongly temperature-dependent below 3K (cf. Fig. S7.) All rows except ``D25 electrons'' refer to the magnetoresistance region on the hole side. For all rows, the resistivity at zero field in the magnetoresistance regime is several tens of Ohms. $B_{sat}$ is a very rough estimate of the lowest field at which the MR is within a few \% of its maximum, MR$_{max}$. These values are hard to quantify to better accuracy than $\sim 10\%$ because quantum oscillations tend to become comparable in amplitude with the magnetoresistance at roughly the field at which it appears to saturate.}
\end{table*}

\setcounter{figure}{0}
\renewcommand\thefigure{\textbf{S\arabic{figure}}}

\newpage
\begin{figure}[t]
\centering
\includegraphics[width=0.5\columnwidth]{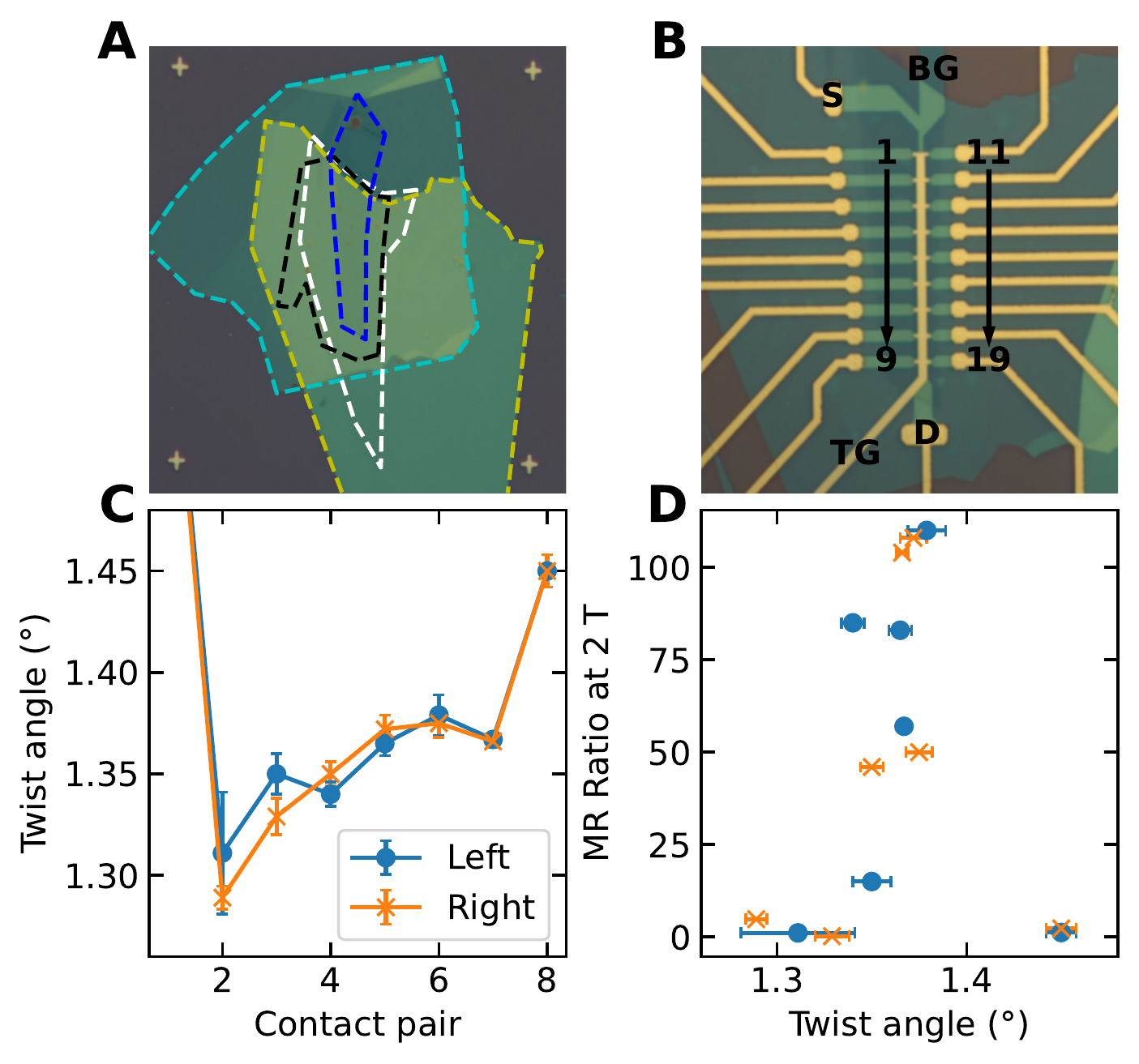}
\captionof{figure}{\textbf{Twist angle variation throughout the device}. (\textbf{A}) Device layout prior to lithography with separate layers outlined. From top to bottom we have top hBN (yellow, 20 nm thick), top graphene (white), bottom graphene (black), bottom hBN (cyan, 13 nm thick), and graphite back gate (blue). The plus-shaped alignment marks in the corners are 100 \textmu m apart. (\textbf{B}) Finished device layout after lithography. The labels are for the current source contact (S), current drain contact (D), back gate (BG), top gate (TG), Hall voltage probes on the left (1-9), and on the right (11-19). (\textbf{C}) Twist angle variation along the device as measured by the longitudinal magnetotransport at high density and field. The first set of contact pairs on either side is at 1.9°. (\textbf{D}) Magnetoresistance ratio at 2 T at fixed density $2.3\times 10^{12}$ cm$^{-2}$ at 1.7 K. The magnetoresistance for the contact pairs at 1.9° (not shown) are 0.8 and 0.3 on the left and right respectively.}
\end{figure}

\begin{figure}[t]
\centering
\includegraphics[width=\columnwidth]{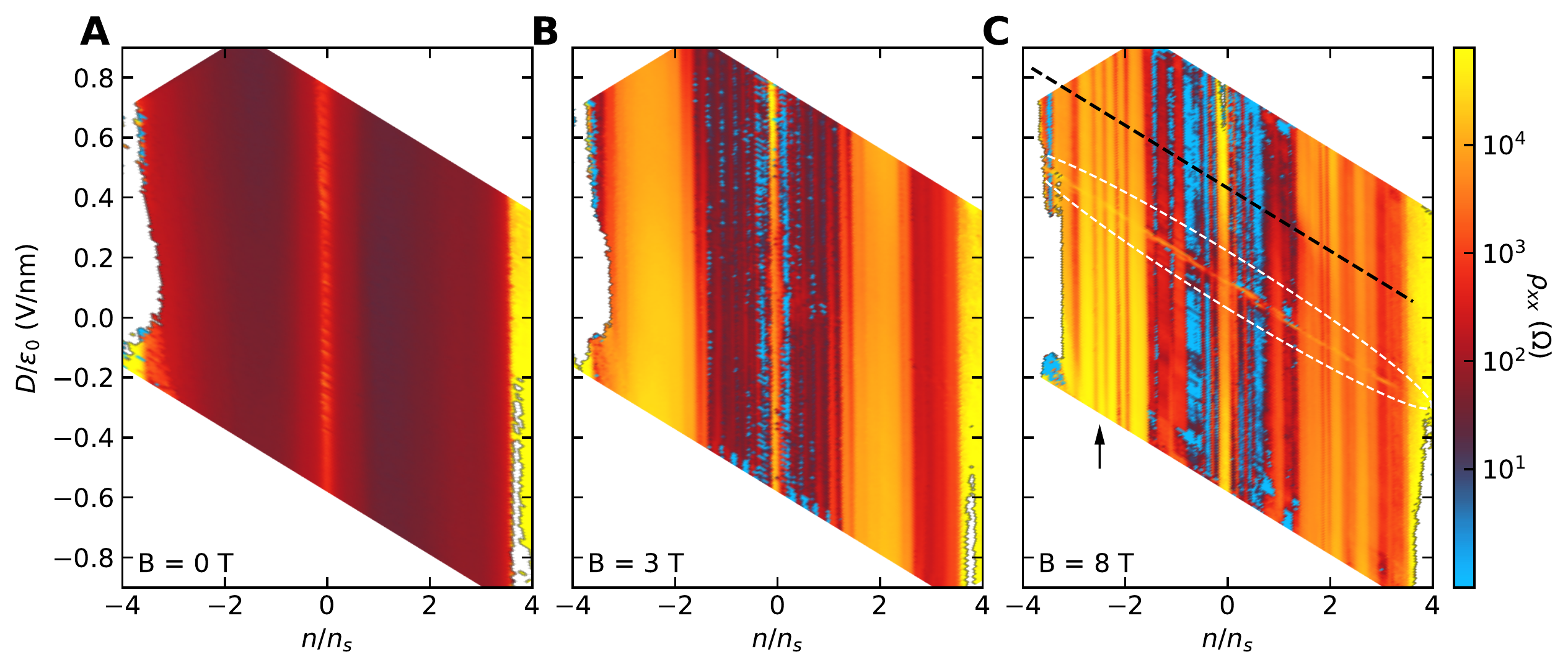}
\captionof{figure}{\textbf{Effect of electric displacement field}. (\textbf{A}) Gate maps at 0 field, (\textbf{B}) at 3 T, and (\textbf{C}) at 8 T. The black dashed line is the cut used for the figures in the main text, while the white ellipse indicates the loss of signal when the back gate is fixed at 0 V. The arrow indicates one pair of split LLs, which appear as pairs of vertical lines within the magnetoresistance regions. That they are vertical indicates that they do not vary with displacement field.}
\end{figure}

\begin{figure}[t]
\centering
\includegraphics[width=\columnwidth]{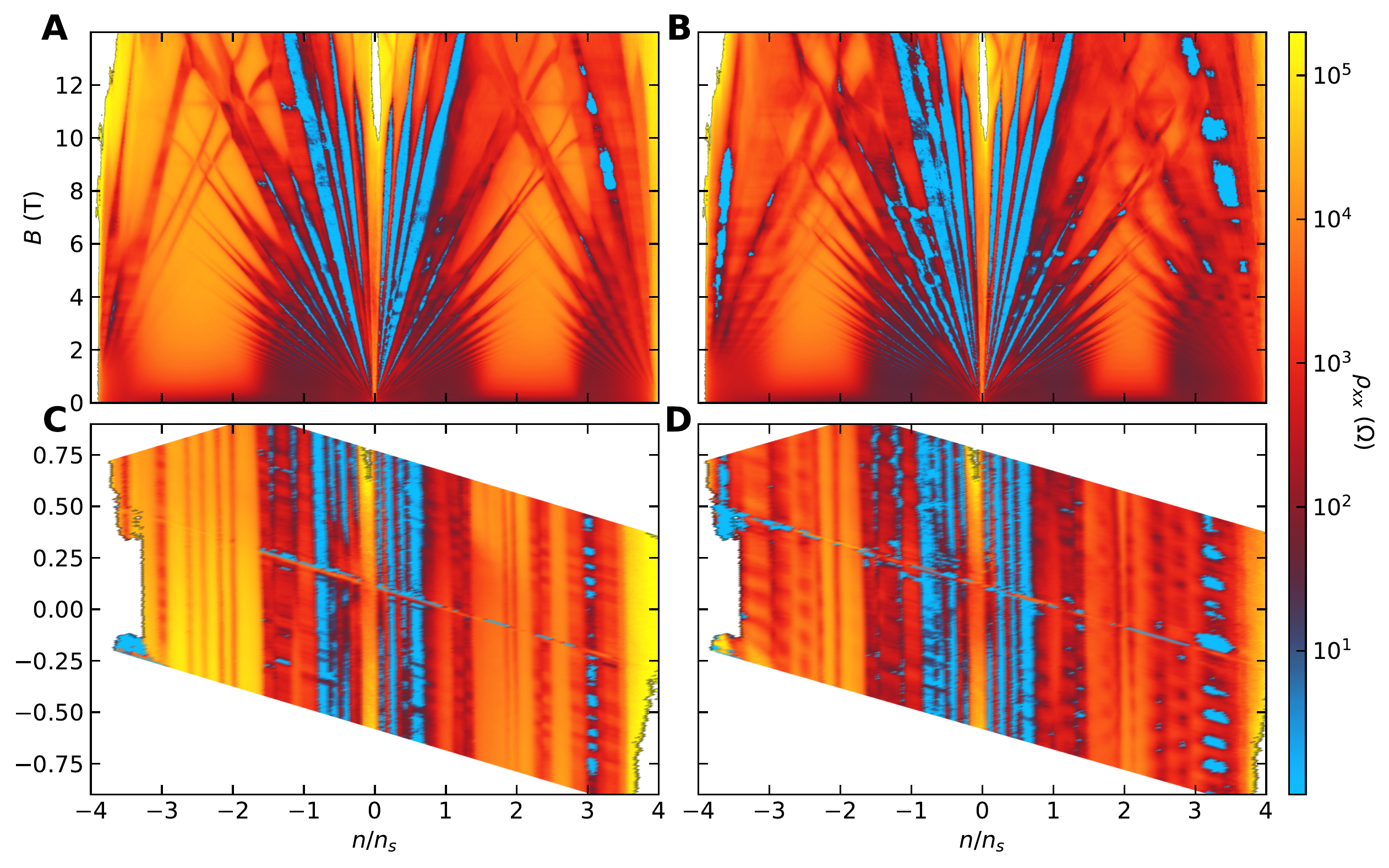}
\captionof{figure}{\textbf{Similar behavior in other contact pairs}. (\textbf{A}) Fan diagram for contact pair 7 - 8 and (\textbf{B}) 4 - 5. (\textbf{C}) Gate map at 8 T for contact pair 7 - 8 and (\textbf{D}) 4 - 5.}
\end{figure}

\begin{figure}[t]
\centering
\includegraphics[width=0.7\columnwidth]{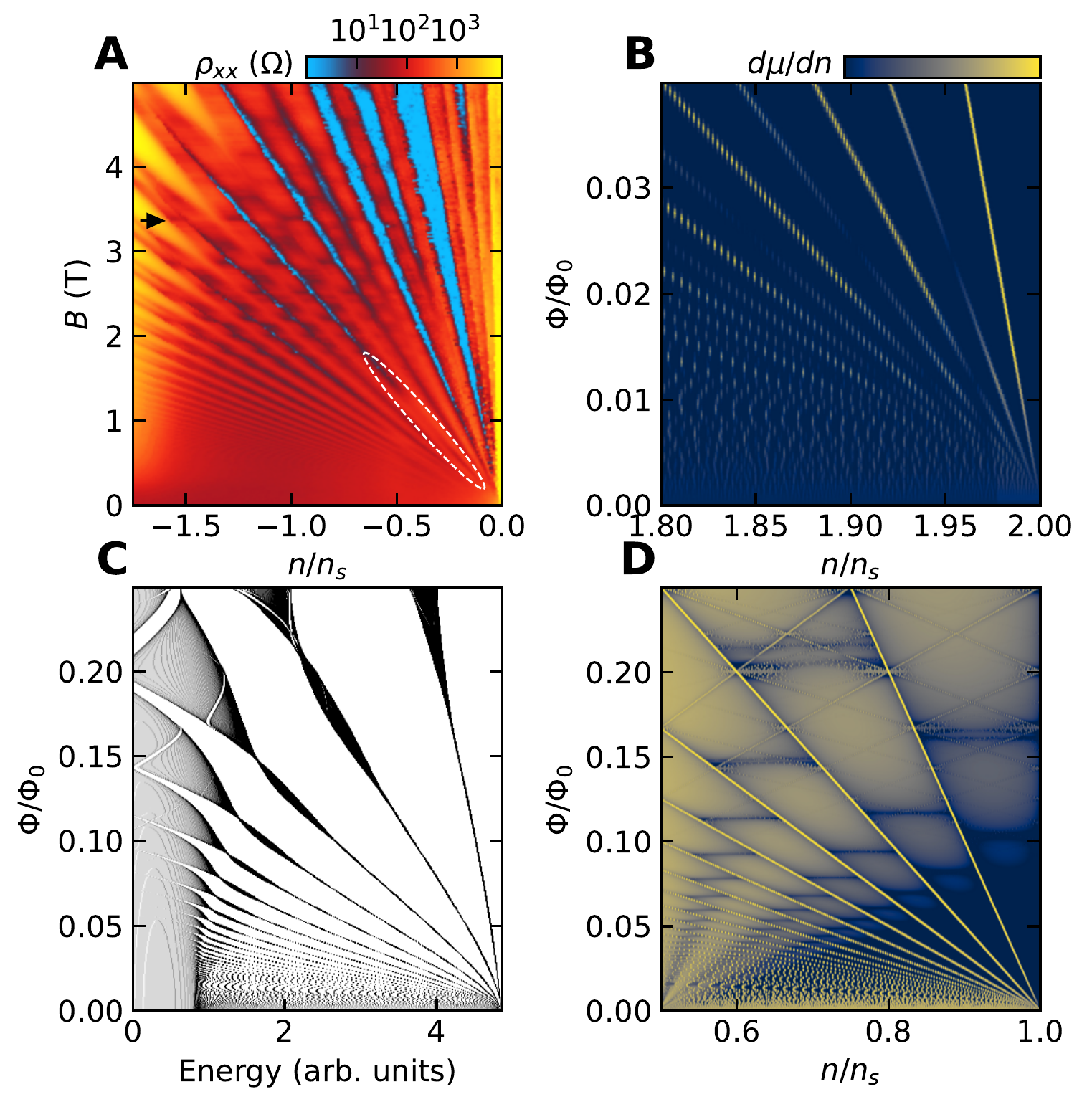}
\captionof{figure}{\textbf{Additional phenomenology captured by the model}. (\textbf{A}) Magnetotransport for contact pair 7 - 8 at low field and low density. The dashed circle indicates the feature reproduced in simulation in panel B, and the arrow indicates one of the features reproduced in C and D. (\textbf{B}) Inverse density of states near a band edge at low field for $a_x=1$, $a_y=2$, $V=0.2$, the same parameters as in Fig 4h. This is taken at $q = 1999$. (\textbf{C}) Energy spectrum for $a_x=2$, $a_y=1$, $a_{2x}=0.6$, where $a_{2x}$ represents hopping two sites in $x$. (\textbf{D}) Inverse density of states for the spectrum in C. In this panel, the color scale is logarithmic.}
\end{figure}

\begin{figure}[t]
\centering
\includegraphics[width=0.5\columnwidth]{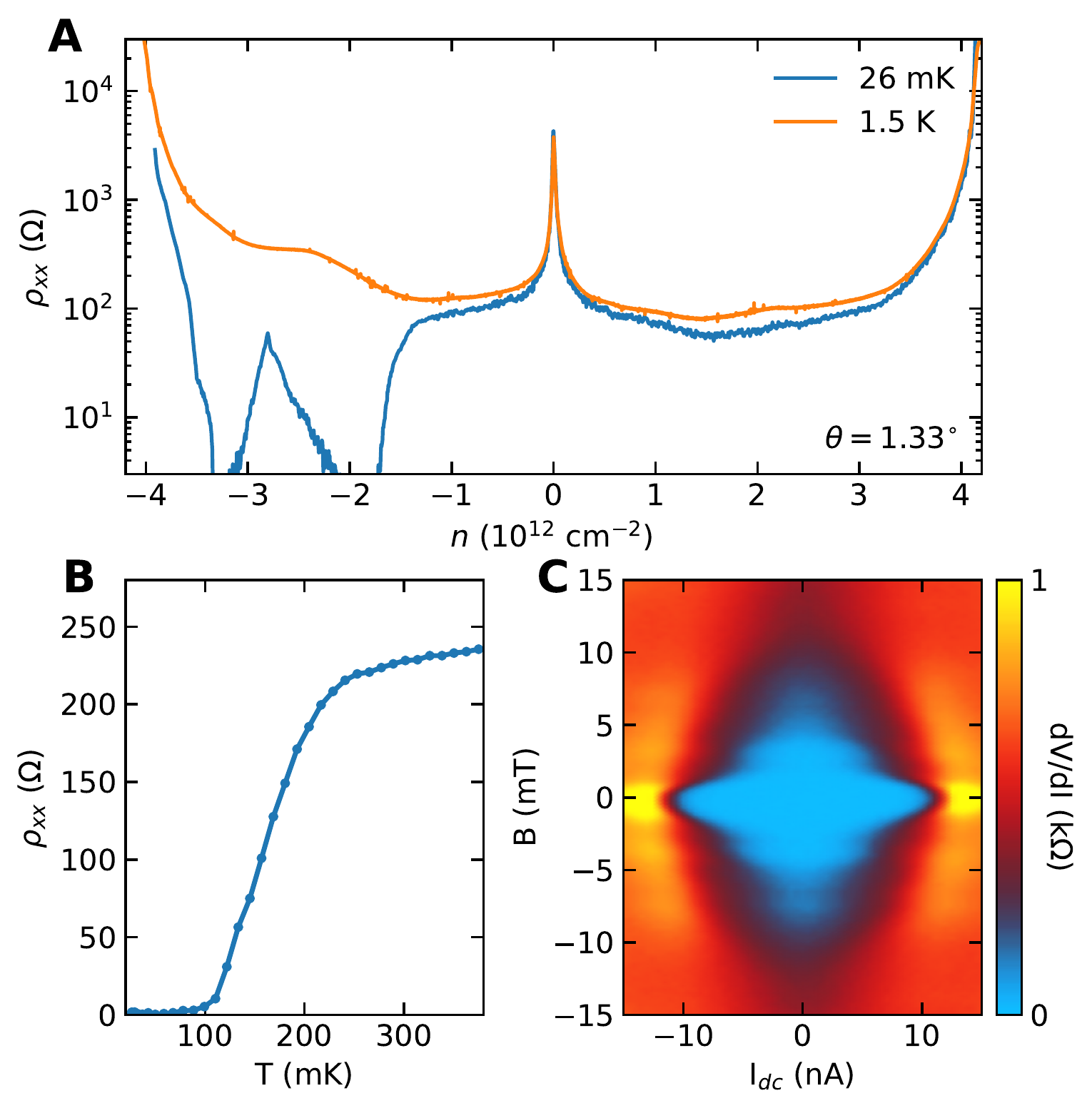}
\captionof{figure}{\textbf{Superconductivity at 1.33°}. (\textbf{A}) Resistivity of contact pair 13 - 14 as a function of density at the indicated temperatures. (\textbf{B}) Temperature dependence of the superconductivity at fixed density at $n=-3.2\times 10^{12}$ cm$^{-2}$. (\textbf{C}) Differential resistance as a function of DC bias current and field demonstrating weak Fraunhofer-like behavior at $n=-3.2\times 10^{12}$ cm$^{-2}$. We have removed a small flux jump around $B = -7$ mT in preparing this figure.}
\end{figure}

\begin{figure}[t]
\centering
\includegraphics[width=0.6\columnwidth]{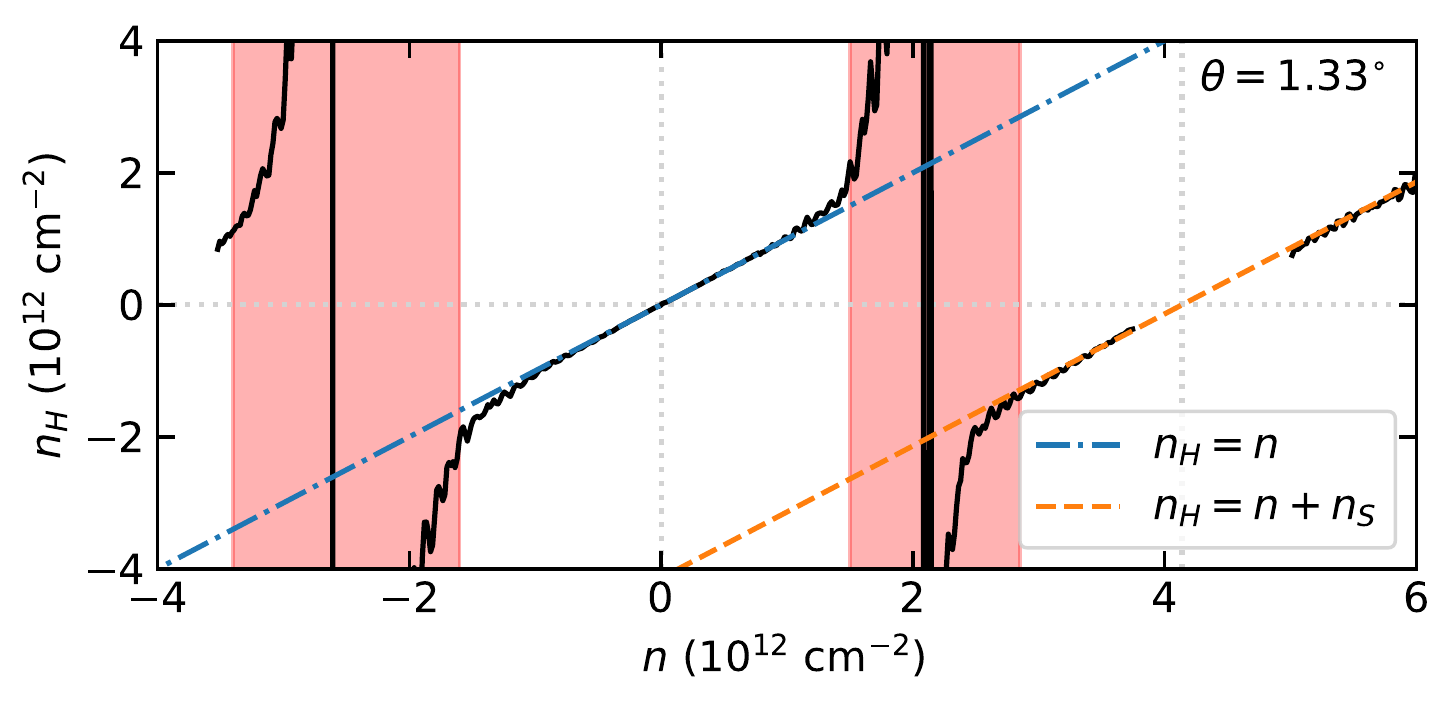}
\captionof{figure}{\textbf{Low-field Hall effect}. The Hall density for contact pair 13 - 3, as measured between -0.5 and 0.5 T. The regions of large magnetoresistance are highlighted in red.}
\end{figure}

\begin{figure}[t]
\centering
\includegraphics[width=0.7\columnwidth]{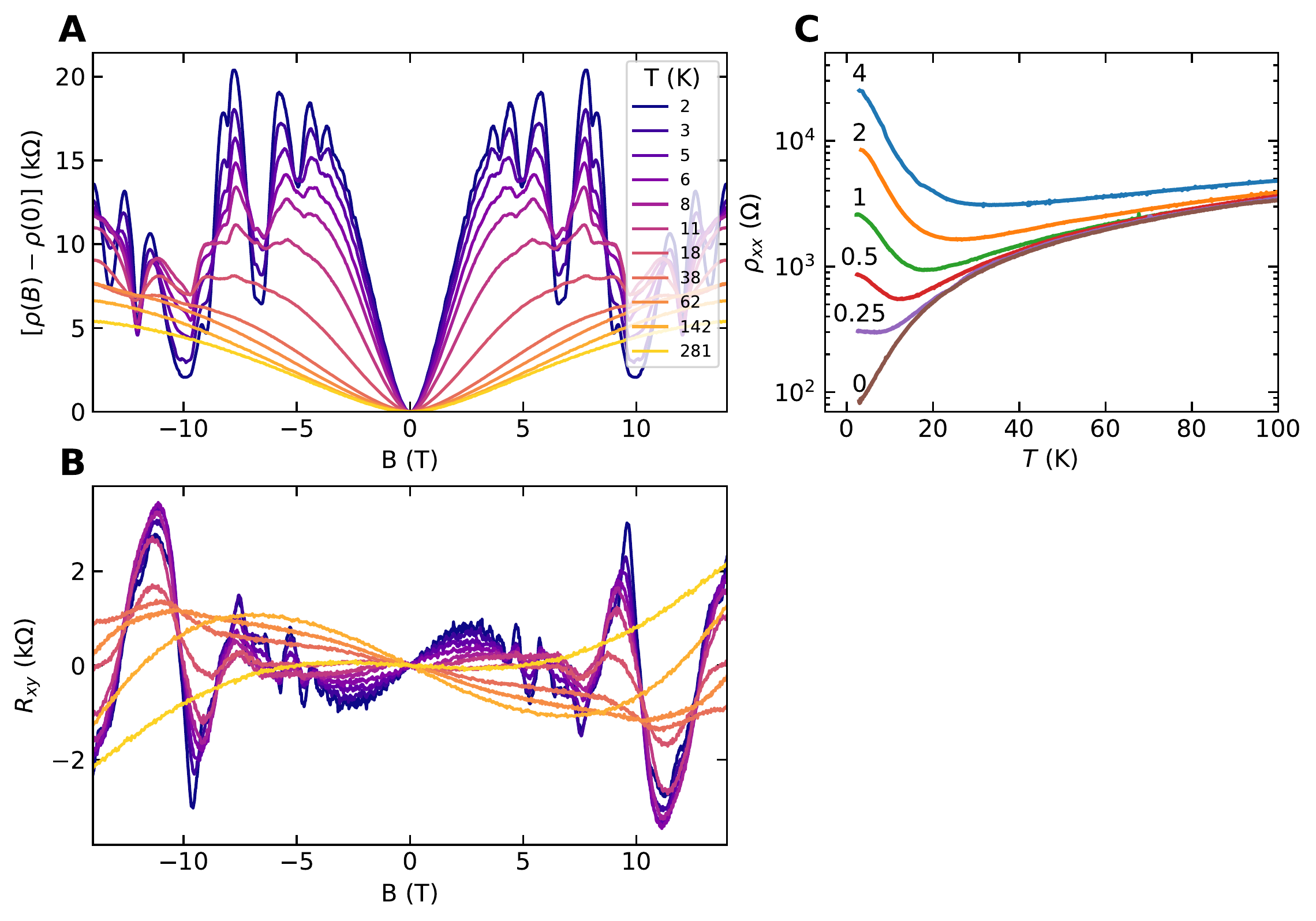}
\captionof{figure}{\textbf{Temperature dependence of large magnetoresistance}. (\textbf{A}) Temperature dependence of longitudinal resistivity (contact pair 16 - 17, symmetrized) and (\textbf{B}) Hall resistance (contact pair 6 - 16, antisymmetrized) at $n/n_s=2.5$. (\textbf{C}) Temperature dependence of longitudinal resistivity at fixed fields at $n/n_s=-2.8$.}
\end{figure}

\begin{figure}[t]
\centering
\includegraphics[width=\columnwidth]{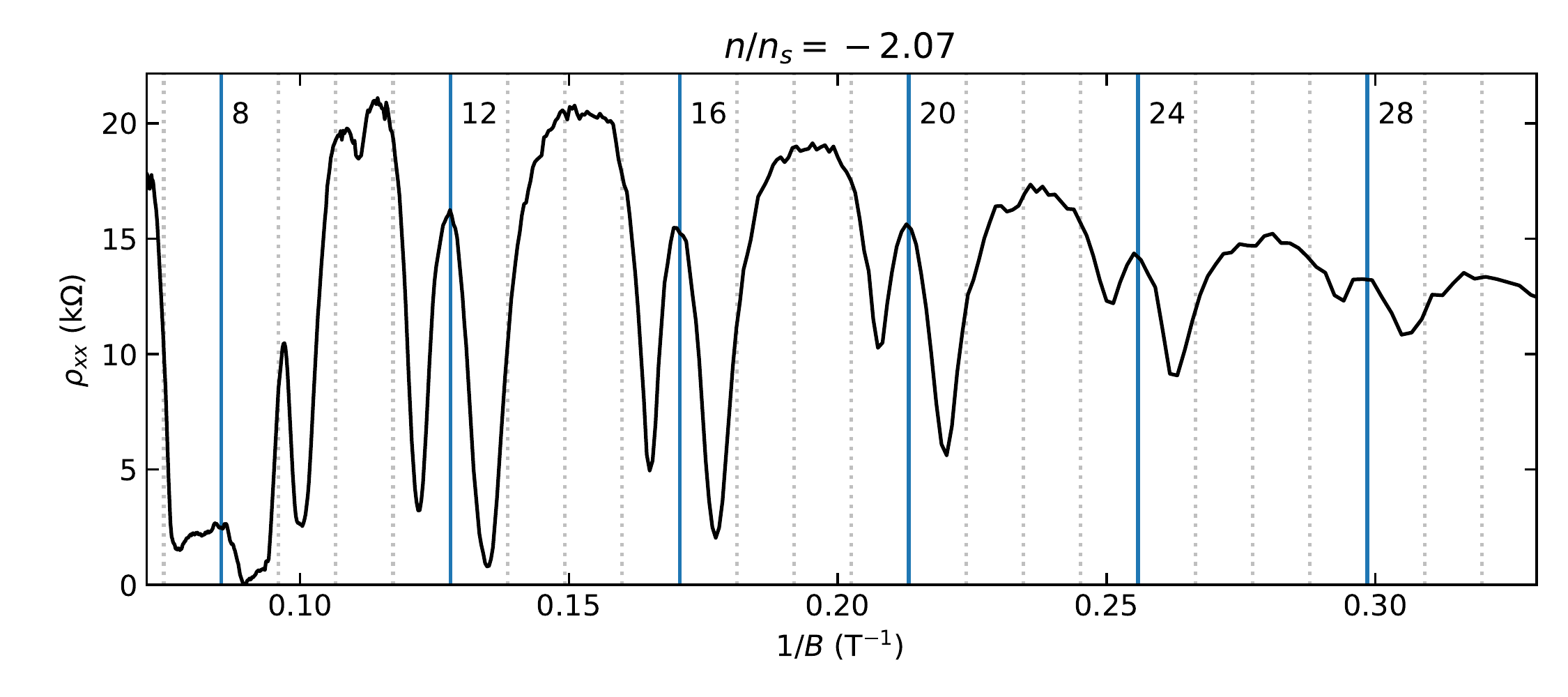}
\captionof{figure}{\textbf{Split Landau levels as a function of $\mathbf{1/B}$}. Longitudinal resistivity at fixed density within the magnetoresistance region as a function of $1/B$. The vertical lines indicate expected Landau level filling factors originating from charge neutrality. Note that the split LLs correspond to roughly $\nu = 11.4, 12.6, 15.4, 16.6, …$ ($\nu = 8$ has extra features originating from $n/n_s = -4$). That the split LLs have the same period in $1/B$ indicates that they correspond to the same density offset from charge neutrality and are not a product of multiple Fermi surfaces or band reorganization.}
\end{figure}

\begin{figure}[t]
\centering
\includegraphics[width=\columnwidth]{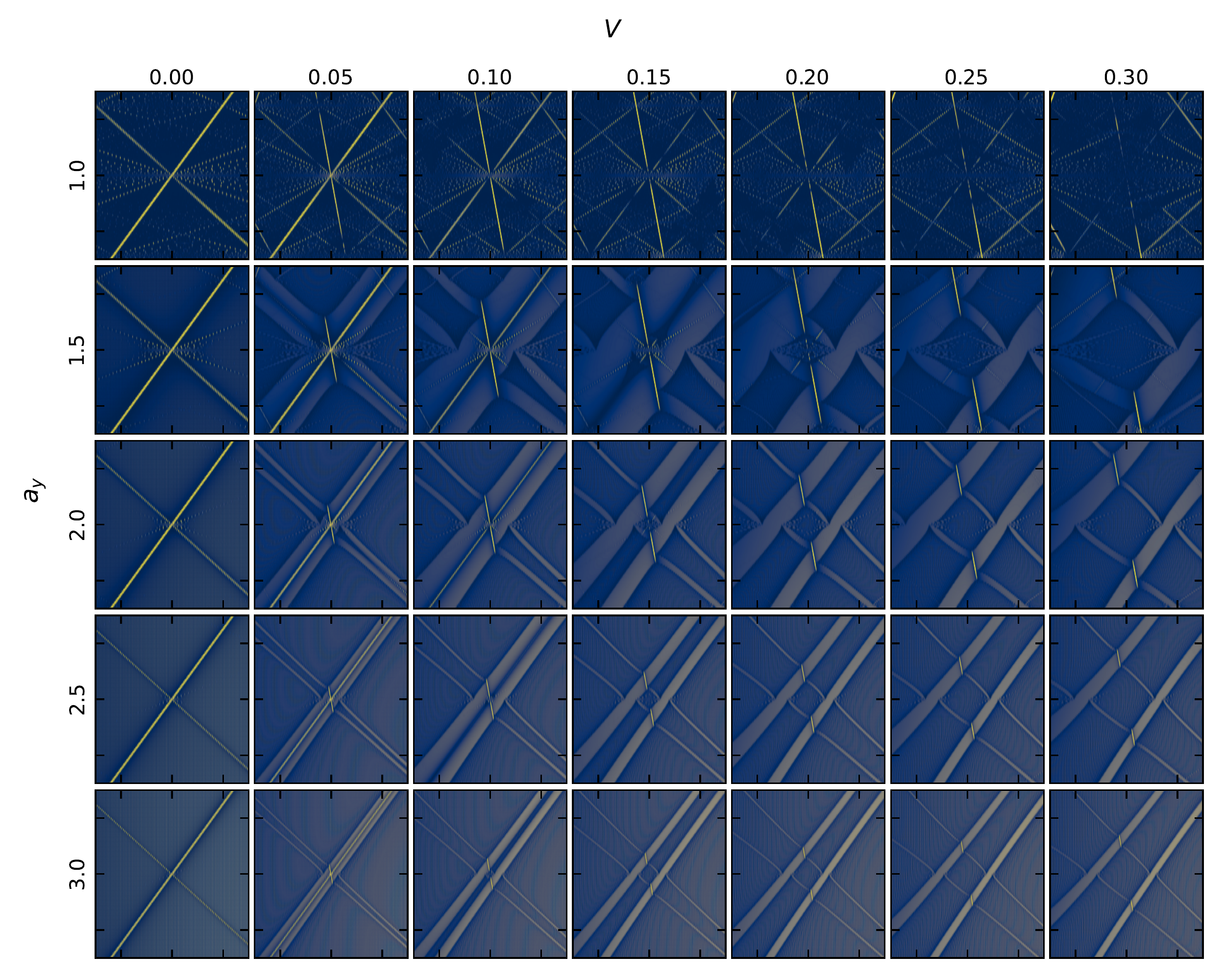}
\captionof{figure}{\textbf{Model behavior as a function of $a_y$ and $V$}. Each plot shows the same region of ($n$, $B$) as in Fig. 3c of the main text, taken at $q = 1999$ and $a_x=1$. The combination of experimentally observed qualitative phenomenology is reproduced by the model not just at a fine-tuned value of the parameters $a_y$ and $V$ but rather for ranges of parameters $a_y$ and $V$ such that $1.5 < a_y < 3$ and $0.1 < V < 0.3$. The ranges are comparable for different ($n$, $B$). For parameter values that show clear split LLs, increasing $V$ increases the distance between the split LLs (which goes to zero when $V = 0$), and increasing $a_y$ decreases their width. Increasing $a_y$ beyond the stated range suppresses the discontinuities at LL crossings, as a direct consequence of the decreasing width of the LLs: compare ($a_y$, $V$) = (2, 0.3) to (3, 0.3).}
\end{figure}

\begin{figure}[t]
\centering
\includegraphics[width=\columnwidth]{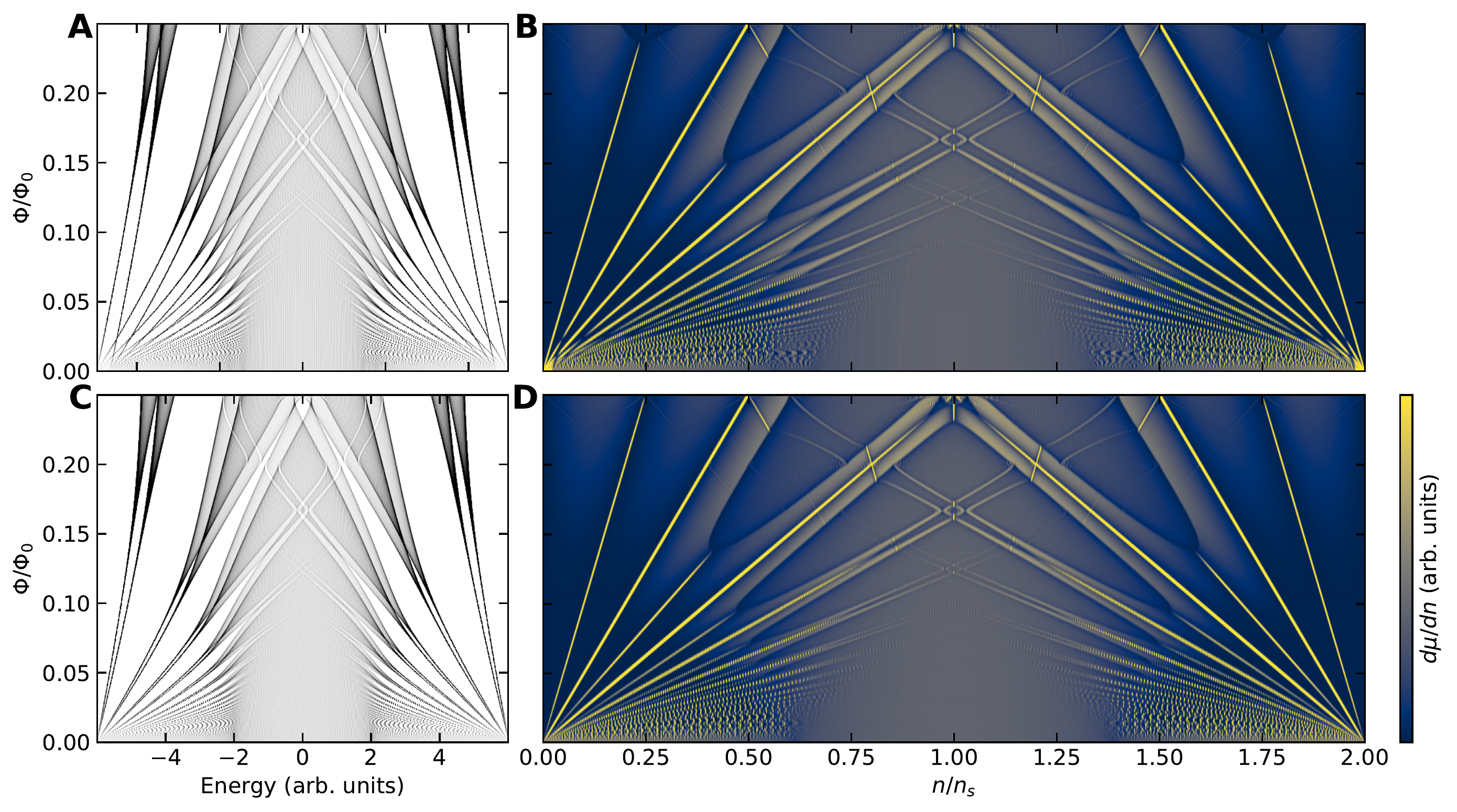}
\captionof{figure}{\textbf{Comparison of Zeeman-like splitting to fixed splitting.} Energy spectra ($q=1499$) and inverse density of states for $a_x=1$, $a_y=2$, and (\textbf{A-B}) $V=0.2$, (\textbf{C-D}) $V=\Phi/\Phi_0$. The two models produce qualitatively very similar phenomenology. There is one notable but subtle differences: At low fields near charge neutrality and full filling, the constant-$V$ Landau levels from the two butterflies cross each other, whereas in the Zeeman-like model the two butterflies are not split at low field and thus cannot cross at low field. These crossings lead in the constant $V$ model to the disappearance and reappearance of gaps along the density/field trajectory of a given Landau level, as in Fig. S4b. This phenomenon seems to happen in experimental plot S13b at ($n=0.9\times 10^{12}$ cm$^{-2}$, $B=3$ T) and less prominently in measurements on at least three contact pairs of the device from the main text that show LL splitting.
Though this may favor the constant $V$ model as a description of the experiment, more measurements would be needed to definitively support that assignment.}
\end{figure}

\begin{figure}[t]
\centering
\includegraphics[width=\columnwidth]{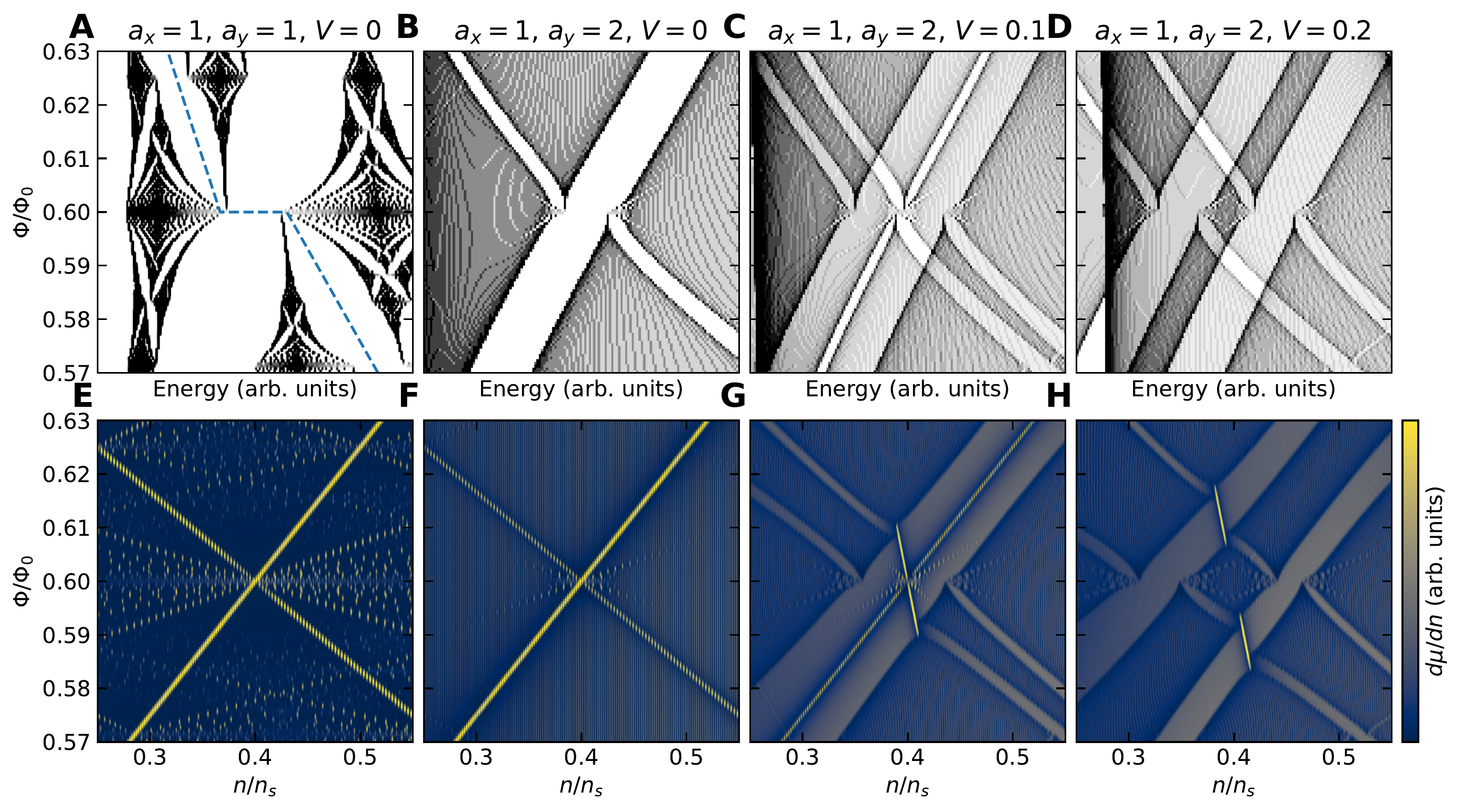}
\captionof{figure}{\textbf{Split LL detail in the model.} (\textbf{A-D}) Energy spectra for the indicated parameters for $q = 1999$, zoomed in to the intersection of the LLs $s, t = 4, -6$ (gap indicated with the dashed line) and $-2, 4$. (\textbf{E-H}) Associated inverse DOS for the respective spectra. Panel \textbf{H} shows the same parameters as Fig. 3c.}
\end{figure}

\begin{figure}[t]
\centering
\includegraphics[width=\columnwidth]{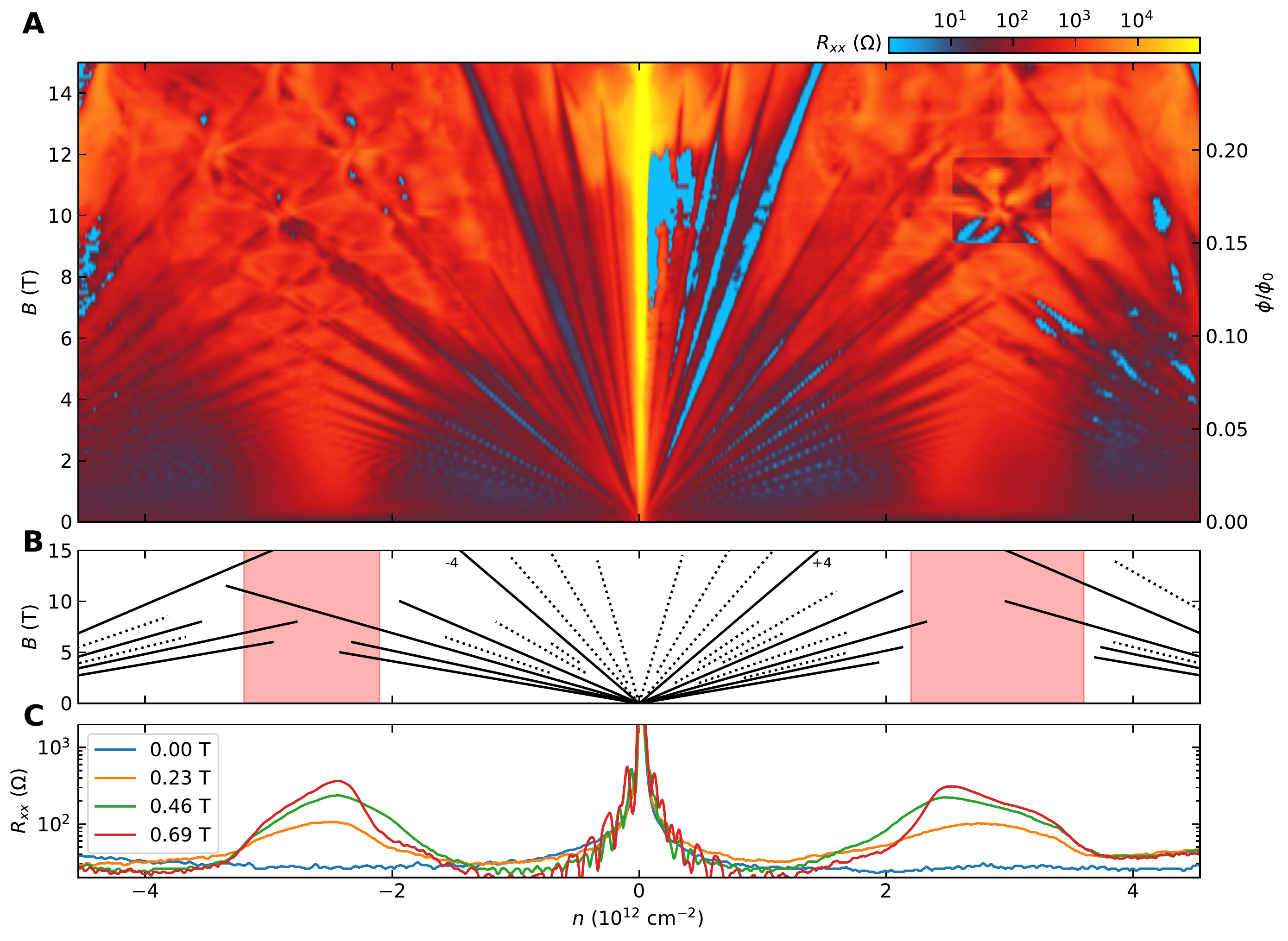}
\captionof{figure}{\textbf{Replication of split LLs in a second device.} (\textbf{A}) Longitudinal resistance vs density and perpendicular magnetic field for device D34 ($\theta = 1.59$°). The window with increased contrast (color bar does not apply) highlights the intersection of $\nu = 12$ from charge neutrality and $\nu = -12$ from full filling. There is a faint vertical line, corresponding to the average of the slopes of the two Landau levels. (\textbf{B}) Schematic fan diagram corresponding to (A), with the same legend as Fig. 2 in the main text. (\textbf{C}) Line cuts from panel A at the indicated fields, showing regions of magnetoresistance with location and shape similar to those in Fig. 1b.}
\end{figure}

\begin{figure}[t]
\centering
\includegraphics[width=\columnwidth]{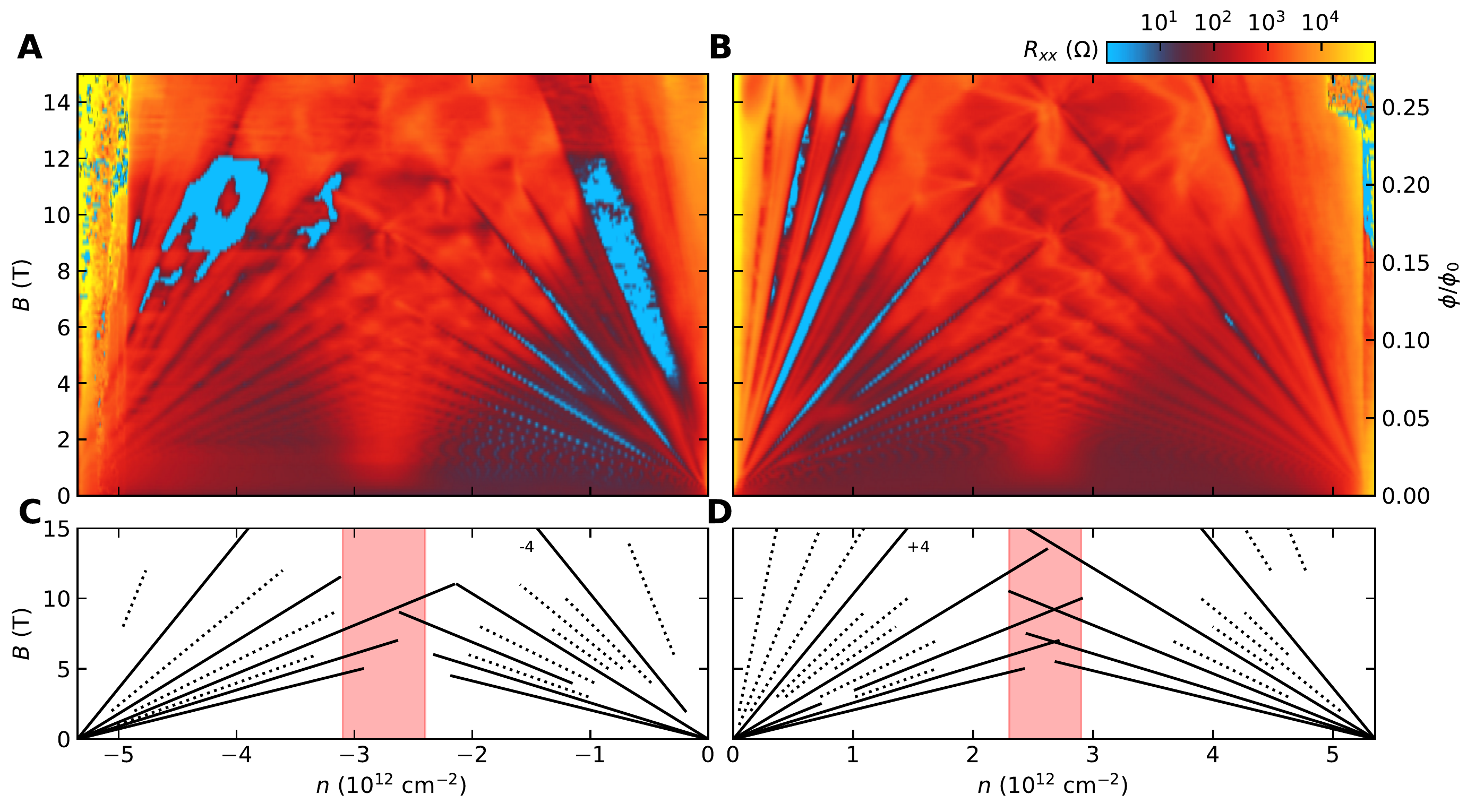}
\captionof{figure}{\textbf{Lack of split LLs in a third device similarly far from magic angle.} Longitudinal resistance vs density and perpendicular magnetic field for device D25 ($\theta = 1.52$°) for holes (\textbf{A}) and electrons (\textbf{B}). The measurements for electrons and holes are taken in different contact pairs and at different Si gate voltages. (\textbf{C-D}) Schematic fan diagram corresponding to (A) and (B) respectively, with the same legend as Fig. 2 in the main text.}
\end{figure}

\end{document}